\documentclass[letterpaper,english,reprint, aps, nofootinbib, prd, twocolumn, lengthcheck, superscriptaddress]{revtex4-1}
\usepackage[T1]{fontenc}
\usepackage[latin9]{inputenc}
\usepackage{verbatim}
\usepackage{bm}
\usepackage{amsmath}
\usepackage{amssymb}
\usepackage{graphicx}
\usepackage{slashed}
\usepackage{wrapfig}
\usepackage{bbm}

\makeatletter

\pdfpageheight\paperheight
\pdfpagewidth\paperwidth

\usepackage[usenames, dvipsnames]{xcolor}
\definecolor{grayflag}{gray}{0.6}

\newcommand{\roma}[1]{\mbox{\footnotesize{#1}}}

\newcommand{\qbarq}{\langle \bar{q}q\rangle}
\newcommand{\qq}{\langle qq \rangle}
\newcommand{\pid}{\tilde{\pi}}  % diquark condensate pion
\newcommand{\sigmad}{\tilde{\sigma}}  % diquark sigma (i.e., delta_s)

\newcommand{\subint}{\mbox{\footnotesize{int}}}
\newcommand{\beq}{\begin{align}}
\newcommand{\eeq}{\end{align}}
\newcommand{\M}{\Sigma}

\usepackage[colorlinks=true,linktocpage=true,linkcolor=blue,citecolor=blue]{hyperref}

\makeatother

\usepackage{babel}

\begin{document}

\title{Generalized Nambu-Goldstone pion in dense matter: a schematic NJL model}

\author{Yifan Song and Gordon Baym}

\affiliation{Department of Physics, University of Illinois at Urbana-Champaign.}

\date{\today}

\begin{abstract}

  Chiral symmetry is always broken in cold, dense matter, by chiral condensation at low densities and by diquark condensation at high density.
We construct here, within a schematic Nambu-Jona-Lasinio (NJL) model, the corresponding generalized Nambu-Goldstone pion, $\pi_G$.
As we show, the $\pi_G$ mode naturally emerges as a linear combination of the $\qbarq$ vacuum pion $\pi$ and the $\qq$ diquark-condensate pion $\pid$, with $q$ the quark field, and continuously evolves with increasing density from being $\pi$-like in the vacuum to  $\pid$-like in the high density diquark pairing phase.  We calculate the density-dependent mass, decay constant, and coupling to quarks of the $\pi_G$, and derive a generalized Gell-Mann--Oakes--Renner (GMOR) relation in the presence of a finite bare quark mass $m_q$.
We briefly discuss the implications of the results to possible Bose condensation of $\pi_G$ in more realistic models.

\end{abstract}

\maketitle

\section{\label{sec:intro}Introduction}

    Chiral symmetry, spontaneously broken in the vacuum and in low density nuclear matter by chiral condensation -- with order parameter $\langle \bar{q}q\rangle$  -- gives rise to the pseudoscalar octet of Nambu-Goldstone (NG) bosons, pions, kaons, and eta.      In the quark matter regime at densities well above normal nuclear matter density,  BCS diquark pairing is predicted \citep{Alford_2008rev}, either in a color-flavor-locked (CFL) phase or a partially paired phase dependent on the density and the $u$, $d$, and $s$ quark masses; the resulting diquark condensates $\langle qq\rangle$ continue to break chiral symmetry at high density\footnote{ Depending on the specific diquark condensation at different densities \citep{Ruster2005, Blaschke2005, Muller2013}, the $SU(3)_L \otimes SU(3)_R$ chiral symmetry may only be partially broken.  In the partially paired ``2SC'' isoscalar phase, likely favored at moderate density where only up and down quarks pair, the isospin subgroup $SU(2)_L \otimes SU(2)_R$ of the chiral symmetry remains unbroken by the 2SC diquark condensate.  On the other hand, in the CFL phase at high density, all eight axial generators of the chiral symmetry are broken by the CFL diquark condensate, in which all quark flavors are paired.       }
     \citep{latticeReview, Fukushima2010rev, Alford_chiSB_CFL, Pisarski1999}, even though the chiral condensates $\langle \bar{q}q\rangle$ gradually disappear.
 As a result, the vacuum meson modes, corresponding to fluctuations of the $\langle \bar{q}q\rangle$ order parameter, become replaced with diquark condensate meson modes corresponding to fluctuations of the $\langle qq\rangle$ order parameter in dense quark matter \citep{dtsonReversal, fukushimaConstruct}.   Between low density nuclear matter and high density quark matter, we expect an extensive coexistence region of finite $\langle \bar{q}q\rangle$ and $\langle qq\rangle$, in which the NG modes are a combination of the vacuum meson modes and diquark-condensate meson modes \citep{ Rho2000,hatsudaCollective, Toru2016}.  
    As the density increases, the chiral NG modes evolve from vacuum mesons to diquark-condensate mesons, and their physical properties such as masses, decay, and interaction with quarks are modified as the $\langle \bar{q}q\rangle $ condensates are gradually replaced by the $\langle qq\rangle$ condensates.

    While chiral NG mesons have been studied in the limits of low density
(non-BCS paired) $\bar{q}q$ condensed matter \citep{Bernard1986, Ebert1995, Buballa2000, He2005, Hansen2007} and high density pure-BCS $qq$ paired matter  \citep{buballa2007,Ruggieri2007,Ebert2008,Rho2000} using the Nambu-Jona-Lasinio (NJL) model \citep{NJL1,NJL2, buballa2005review},  a quantitative description of NG mesons at intermediate  densities remains an open problem.  Such a description requires adopting specific models to describe the changing phase structure with increasing density, itself an unresolved issue \citep{Fukushima2010rev}.     In this paper, we  study the chiral structure of a simplified single flavor, single color  NJL model that includes both scalar and pseudoscalar condensates.  Such a model has a single chiral NG mode, which we refer to as the {\em generalized pion},\footnote{The name ``generalized mesons'' was used, e.g., in \citep{hatsudaCollective}, to describe the $\bar{q} \bar{q} qq$ modes corresponding to fluctuations of the diquark condensates at high density.  For clarity, we refer in this paper to the NG modes (a combination of $\bar{q}q$ and $\bar{q} \bar{q} qq$ modes) as ``generalized mesons,"  the $\bar{q} \bar{q} qq$ modes as ``diquark-condensate mesons,'' and the usual $\bar{q}q$ modes as ``vacuum mesons."} $\pi_G$,  corresponding to simultaneous fluctuations of the 
$\langle \bar{q}q\rangle$ and $\langle qq\rangle$ order parameters.   The resulting phase diagram, with properly chosen model interaction parameters, mimics the more realistic QCD phase diagram in terms of chiral symmetry breaking by the low and high density condensates, which are here connected smoothly by a coexistence region (for sophisticated NJL constructions of QCD phase diagram, see e.g., \citep{Blaschke2005, Hatsuda1994, Fukushima2004, Warringa2005, Ratti2006, Ratti2007, Abuki2008, Abuki2010, Shao2011, Schwarz1999, powell2012, powell2013, philipthesis}).  The generalized pion continuously evolves from  the vacuum pion, $\pi$, in the low density chirally broken phase to the diquark-condensate pion, $\pid$, in the high density BCS phase; its mass and decay constant are continuous functions of quark density, and obey a generalized Gell-Mann--Oakes--Renner (GMOR) relation, which we calculate to second order in $m_q$.  Its coupling vertex to the quark field also changes continuously with increasing density.  

   The present study is a first step in understanding in detail the density-dependent properties of the pseudoscalar mesons extrapolated into high density quark matter, and is readily generalized to more realistic models with multiple flavors and colors to quantitatively study the meson mass ordering reversal problem \citep{dtsonReversal}.   In addition to clarifying the QCD phase diagram in terms of generalized meson condensation, the study of the $\pi_G$ mode also contributes to understanding the thermodynamics of dense matter, and thus eventually the interiors and cooling of neutron stars \citep{Yakovlev2001}.

   In Sec.~\ref{subsec:Lagrangian} of this paper we introduce
the model NJL Lagrangian, analytically solve it in the mean field approximation, and discuss the quasiparticle spectrum and energy eigenvectors, while in Sec.~\ref{sec:scalar} we compute the quark propagator as well as the gap equations in the even parity, spin-singlet, or ``scalar,'' ground state, without pion condensation.   Then, in Sec.~\ref{subsec:Stability} we investigate the phase diagram and thermodynamic stability of the system as a function of the model chiral and diquark coupling strengths, which enables us to restrict the parameter space in terms of an ultraviolet cutoff, in order that the resulting phase diagram includes a chirally broken vacuum phase and
a high density BCS phase, connected by a coexistence phase at intermediate density, thus mimicking the more realistic phase diagrams in NJL studies of cold dense matter.

    We next discuss the collective modes in detail in Sec.~\ref{subsec:Collective_modes}.
We first identify all the six collective modes in the chiral limit in Sec.~\ref{subsec:mode_para} corresponding to fluctuations of the chiral and diquark order parameters $\langle \bar{q}q\rangle$ and $\langle qq\rangle$.  We then focus on the two pseudoscalar pionic modes $\pi$ and $\pid$ in particular, calculating their mixing mass matrix in Sec.~\ref{subsec:mode_mass} and their decay constants in Sec.~\ref{subsec:mode_decay}, relating them to the mass and decay constant of the re-diagonalized NG mode $\pi_G$, and we then derive the density-dependent coupling vertex of $\pi_G$ to quarks in the medium.
    In Sec.~\ref{subsec:mode_mq} we look at the modifications introduced by a finite bare quark mass $m_q$, e.g., its effect on the $\pi_G$ mass.  We derive the matrix generalization of the GMOR relation, deriving the two masses of the two pionic modes to second order in $m_q$, and discuss their behavior with varying density.  Finally, in Sec.~\ref{sec:Outlook} we briefly comment on the implications of possible condensates of the NG mode in quark matter, together with several other open questions, such as the possible roles of a new massive mode corresponding to the phase difference between scalar and pseudoscalar diquark condensates.

Throughout we assume zero temperature unless stated otherwise, and use units $\hbar = c = 1$.

\section{\label{subsec:Lagrangian}Lagrangian, gap equations, quasiparticle dispersion relations and energy eigenstates}

We focus on the Lagrangian,
\begin{eqnarray}
\mbox{\ensuremath{\mathcal{L}}} & = & \bar{q}\left(i\slashed{\partial}-m_{q}+\gamma_{0}\mu\right)q+G\left[\left(\bar{q}q\right)^{2}+\left(\bar{q}i\gamma_{5}q\right)^{2}\right]\label{eq:Lagrangian0}\\
 &  & +H\left[\left(q^{T}i\gamma_{5}Cq\right)\left(\bar{q}i\gamma_{5}C\bar{q}^{T}\right)+\left(q^{T}Cq\right)\left(\bar{q}C\bar{q}^{T}\right)\right],\nonumber 
\end{eqnarray}
where $q$ is the quark field with bare mass $m_{q}$ and quark chemical
potential $\mu$; $\gamma^\mu$ are Dirac gamma matrices and $\gamma^5 = i\gamma^0\gamma^1\gamma^2\gamma^3$,$C=i\gamma^{0}\gamma^{2}$ is the charge conjugate
matrix, $G$ is the coupling strength for the four-quark chiral interaction
term, and $H$ is the strength of the spin-singlet pairing interaction;
$G$ and $H$ are model parameters. The 4-quark interaction terms
exhibit equal coupling in the scalar and pseudoscalar channels. As
a result, the model for vanishing $m_{q}$ has a $U(1)_L\otimes U(1)_R$ chiral
symmetry\footnote{In our single flavor schematic model we call the $U(1)_L\otimes U(1)_R = U(1)_V\otimes U(1)_A$ symmetry simply the {\em chiral symmetry}, in contrast to realistic NJL models with $N_f >1$, where the $SU(N_f)_L\otimes SU(N_f)_R$ symmetry is the ``chiral symmetry'' and $U(1)_A$ is the $U(1)$ axial symmetry.}  of the
quark field. Finally, as the four-fermion interaction in this model
is not renormalizable, we will adopt a three-momentum cutoff $\Lambda$
to regulate the momentum integrals throughout this work.

We solve this model in the mean field approximation. We define the
vacuum expectation value of the composite operators,\footnote{Our definition of diquark pairing amplitude has $i\gamma_{5}$ between
the quark fields, compared with $\gamma_{5}$ alone, which is often used,
e.g., in \citep{hatsudaCollective}.}
\begin{align}
\sigma & =2G\langle\bar{q}q\rangle,\ \pi=2G\langle\bar{q}i\gamma_{5}q\rangle,\nonumber \\
\Delta_{s} & =2H\langle\bar{q}i\gamma_{5}C\bar{q}^{T}\rangle,\ \Delta_{ps}=2H\langle\bar{q}C\bar{q}^{T}\rangle;\label{eq:MFdef}
\end{align}
here $\Delta_{s}$ is the pairing amplitude in scalar channel, and
$\Delta_{ps}$ in pseudoscalar channel. The condensates $\sigma$
and $\pi$ serve as the order parameters of spontaneous chiral symmetry
breaking at low density;  the fluctuations of $\sigma$ and $\pi$ around their ground state values correspond to the NG boson (the vacuum pion) and the massive Higgs-like mode.   Also, under axial $U(1)_{A}$
rotation $\Delta_{s}$ and $\Delta_{ps}$ rotate into each
other; thus, a non-vanishing expectation value of either of these diquark operators
also indicates broken chiral symmetry. We work in the homogeneous
phase, so that the mean fields are constant in space. 

We use the Nambu-Gor'kov formalism, defining the charge conjugate
quark field $q^{C}=C\bar{q}^{T}$, and forming the Nambu-Gor'kov spinor
$\psi\equiv (q, q^{C})^{T}/\sqrt{2}$. Keeping leading order
fluctuations of the composite operators around their expectation values,
we arrive at the mean field Lagrangian, 
\begin{equation}
\mathcal{L}_{MF}=\bar{\psi}S^{-1}_{MF}\psi-\frac{\sigma{}^{2}+\pi{}^{2}}{4G}-\frac{|\Delta_{s}|^{2}+|\Delta_{ps}|^{2}}{4H},\label{eq:Lagrangian_MF}
\end{equation}    
with the fermion inverse propagator
\begin{equation}
S^{-1}_{MF}=\left(\begin{array}{cc}
i\slashed{\partial}-\hat{M}+\gamma^{0}\mu & \ i\gamma_{5}\Delta_{s}^{*}+\Delta_{ps}^{*}\\
i\gamma_{5}\Delta_{s}+\Delta_{ps} & \ i\slashed{\partial}-\hat{M}-\gamma^{0}\mu
\end{array}\right),\label{eq:InvProp_all}
\end{equation}   
where the effective quark mass matrix is $\hat{M}=m_{q}-\sigma-i\gamma_{5}\pi$.
The quark eigenstates are quasiparticles of momentum $\bm{p}$ with
dispersion relation $\omega(\bm{p})$, given by the solution of
\begin{equation}
\det S^{-1}_{MF}(\omega(\bm{p}),\bm{p})=0\label{eq:detInvProp}
\end{equation}
in frequency-momentum space. The result is
\begin{align}
 & \omega(\bm{p})=\pm\left[(m_{q}-\sigma)^{2}+\pi^{2}+\bm{p}^{2}+\mu^{2}\right.\nonumber \\
 & \hspace{1.2cm}\left.+|\Delta_{s}|^{2}+|\Delta_{ps}|^{2}\pm2\delta(\bm{p})\right]^{\frac{1}{2}};\nonumber \\
 & \delta(\bm{p})\equiv\left[\left(|\bm{p}|\mu\pm\mbox{Im}\left[\Delta_{s}\Delta_{ps}^{*}\right]\right)^{2}\right.\nonumber \\
 & +\left.\mu^{2}\left((m_{q}-\sigma)^{2}+\pi^{2}\right)+\left|(m_{q}-\sigma)\Delta_{ps}-\pi\Delta_{s}\right|^{2}\right]^{\frac{1}{2}}.\label{eq:omega_full}
\end{align}

The leading ``$\pm$'' sign in $\omega(\bm{p})$ is the degeneracy
introduced by the Nambu-Gor'kov formalism; the second ``$\pm$'' sign
in front of $\delta(\bm{p})$ corresponds to the particle-hole branches;
and the last ``$\pm$'' sign within $\delta(\bm{p})$ is a splitting
caused by a relative phase between $\Delta_{s}$ and $\Delta_{ps}$.
All three ``$\pm$'' signs are independent of each other, making a
total of eight eigenvalues (or four physical ones, after removing
the Nambu-Gor'kov degeneracy and keeping only positive $\omega(\bm{p})$).
In the chiral limit $m_{q}=0$, all four of the combinations
\begin{equation}
|\Delta_{s}|^{2}+|\Delta_{ps}|^{2},\,\mbox{Im}\left(\Delta_{s}\Delta_{ps}^{*}\right),\,\sigma^{2}+\pi^{2},\,\mbox{and}\,|\sigma\Delta_{ps}+\pi\Delta_{s}|^{2}\label{eq: 4inv}
\end{equation}
are fully invariant under $U(1)_L\otimes U(1)_R$ rotations, so
that $\omega(\bm{p})$ is always invariant under the full symmetry
group of the Lagrangian. 

In terms of the different quasiparticle eigenvalues $\omega(\bm{p})$,
the grand thermodynamic potential per unit volume $\Omega(T,\mu)$ is given by
\begin{align}
\Omega= & -T\sum_{i=1}^{4}\int_{\bm{p}}\left[\ln\left(1+e^{-\omega_{i}/T}\right)+\frac{1}{2T}\left(\omega_{i}-\omega_{i0}\right)\right]\nonumber \\
 & +\frac{\sigma{}^{2}+\pi{}^{2}}{4G}+\frac{|\Delta_{s}|^{2}+|\Delta_{ps}|^{2}}{4H}.\label{eq:FE_full}
\end{align}
We use $\int_{\bm{p}}$ to denote $\int d^{3}p/(2\pi)^{3}$,
and the summation is over the four positive eigenvalues $\omega_{i}$.
The eigenvalues $\omega_{i0}$ are given by the $\omega_{i}$ with
the mean fields set equal to zero and $\mu =0$, and thus the free energy $\Omega$ vanishes in the vacuum with no condensates. The value of the mean
fields are self-consistently determined by minimizing $\Omega$, resulting
in a total of six equations:
\begin{equation}
\frac{\partial\Omega}{\partial\sigma}=\frac{\partial\Omega}{\partial\pi}=\frac{\partial\Omega}{\partial\Delta_{s}}=\frac{\partial\Omega}{\partial\Delta_{s}^{*}}=\frac{\partial\Omega}{\partial\Delta_{ps}}=\frac{\partial\Omega}{\partial\Delta_{ps}^{*}}=0,\label{eq:GE_derivative}
\end{equation}
which we simply refer to as ``gap equations.'' Only five are
independent; they determine the two chiral fields and the two complex
pairing gaps (to within an overall phase).  In the chiral limit, only
four of the gap equations are independent.

\section{\label{sec:scalar}Scalar chiral and diquark condensates}

Owing to the symmetries of the Lagrangian (\ref{eq:Lagrangian0}), the
solutions to the gap equations (\ref{eq:GE_derivative}) are degenerate.
We focus on the particular choice in chiral limit $m_{q}=0$:
\begin{equation}
\sigma=-M,\ \pi=0,\ \Delta_{s}=-i\Delta,\ \Delta_{ps}=0.\label{eq:normalstate_def}
\end{equation}
This ``scalar state'' describes an even-parity, spin-singlet ground
state without pion condensation. In this state, the NG boson
corresponding to chiral symmetry breaking originates from fluctuations
in $\pi$ and $\Delta_{ps}$, which are pseudoscalar. There are two
reasons for the choice\,(\ref{eq:normalstate_def}):  as in realistic
chiral symmetry breaking in QCD, the NG boson for chiral symmetry
breaking is pseudoscalar. In addition the favored diquark pairing
channel in ground state at high density is likely to be scalar \citep{Alford_2008rev}.
Therefore, we focus on the quasiparticle properties and collective
modes of this particular state.

The quark inverse propagator in the scalar state takes the form
\begin{equation}
S_{0}^{-1}(\omega,\bm{p})=\left(\begin{array}{cc}
\slashed{p}-M+\gamma_{0}\mu & -\gamma_{5}\Delta\\
\gamma_{5}\Delta & \slashed{p}-M-\gamma_{0}\mu
\end{array}\right).\label{eq:InvProp_normalstate}
\end{equation}
The effective mass $M$ and BCS gap $\Delta$ are real. By choosing
$\Delta_{s}$ to be purely imaginary as in Eq.~(\ref{eq:normalstate_def}), the eigenvectors of the Hamiltonian
can be chosen to be entirely real. In the scalar state the two distinct
positive eigenvalues have the familiar form:
\begin{equation}
\omega_{\pm}(\bm{p})=\sqrt{(\epsilon_{\pm}(\bm{p})-\mu)^{2}+\Delta^{2}},\ \epsilon_{\pm}(\bm{p})=\pm\sqrt{\bm{p}^{2}+M^{2}},\label{eq:omega_normalstate}
\end{equation}
each with spin degeneracy two, giving four positive eigenvalues in
total. The corresponding normalized eigenvectors are 
\begin{align}
 & \lambda_{\pm}(\omega_{\pm}(\bm{p}),s)=R_{\pm}(\bm{p})\left(\begin{array}{c}
v_{\pm}(\bm{p})r(\bm{p})\\
u_{\pm}(\bm{p})t(\bm{p})
\end{array}\right),\nonumber \\
 & r(\bm{p})\equiv\left(\begin{array}{c}
s\\
\hat{P}s
\end{array}\right),\ t(\bm{p})\equiv\left(\begin{array}{c}
-\hat{P}s\\
s
\end{array}\right),\label{eq:eigenvectorP_normalstate}
\end{align}
where $s=(1,0)^{T}$ or $(0,1)^{T}$ are spin-1/2 spinors, $R^2_{\pm}(\bm{p})\equiv(\epsilon_{\pm}(\bm{p})+M)/2\epsilon_{\pm}(\bm{p})$
defines the normalization constant, and $\hat{P}\equiv\bm{\sigma}\cdot\bm{p}/(\epsilon_{\pm}(\bm{p})+M)$
is a projection operator in spinor space; and the coherence functions
$v_{\pm}(\bm{p}),u_{\pm}(\bm{p})$ are exactly analogous to the non-relativistic
BCS results:
\begin{align}
v_{\pm}(\bm{p}) & =\sqrt{\frac{\omega_{\pm}(\bm{p})+\epsilon_{\pm}(\bm{p})-\mu}{2\omega_{\pm}(\bm{p})}},\nonumber \\
u_{\pm}(\bm{p}) & =\sqrt{\frac{\omega_{\pm}(\bm{p})-\epsilon_{\pm}(\bm{p})+\mu}{2\omega_{\pm}(\bm{p})}};\label{eq:UV_normalstate}
\end{align}
they satisfy
\begin{equation}
v_{\pm}(\bm{p})^{2}+u_{\pm}(\bm{p})^{2}=1;\ v_{\pm}(\bm{p})u_{\pm}(\bm{p})=\frac{\Delta}{2\omega_{\pm}(\bm{p})}.\label{eq:UV_identies}
\end{equation}
The eigenvectors corresponding to the remaining four negative eigenvalues,
coming from the charge conjugate fields, are instead
\begin{equation}
\tilde{\lambda}_{\pm}(-\omega_{\pm}(\bm{p}),s)=\left(\begin{array}{c}
u_{\pm}(\bm{p})r(\bm{p})\\
-v_{\pm}(\bm{p})t(\bm{p})
\end{array}\right),\label{eq:eigenvectorH_normalstate}
\end{equation}
In our notation, ``$+$'' corresponds to the particle-antihole branch,
and ``$-$'' to the hole-antiparticle branch.

With these explicit eigenvalues and eigenvectors, the quark propagator can be written as
\begin{align}
S_0(\omega,\bm{p})= & \sum_{\pm,s}\left[\lambda_{\pm}(\omega_{\pm}(\bm{p}),s)\lambda_{\pm}^{\dagger}(\omega_{\pm}(\bm{p}),s)\frac{1}{\omega-\omega_{\pm}(\bm{p})}\right.\nonumber \\
 & \left.+\tilde{\lambda}_{\pm}(-\omega_{\pm}(\bm{p}),s)\tilde{\lambda}_{\pm}^{\dagger}(-\omega_{\pm}(\bm{p}),s)\frac{1}{\omega+\omega_{\pm}(\bm{p})}\right]\gamma_{0},\label{eq:Prop_normalstate}
\end{align}
where all eight eigenvalues are summed over. This form is useful for
computing various correlation functions. Lastly, the gap equations
(\ref{eq:GE_derivative}) in the scalar phase reduce to the two independent
equations:
\begin{align}
\frac{M}{2G} & = M\sum_{\pm}\int_{\bm{p}}\frac{1}{\omega_{\pm}(\bm{p})}\left(1\mp\frac{\mu}{\sqrt{\bm{p}^{2}+M^{2}}}\right),\label{eq:GE_m_normalstate}\\
\frac{\Delta}{2H} & = \Delta\sum_{\pm}\int_{\bm{p}}\frac{1}{\omega_{\pm}(\bm{p})}.\label{eq:GE_Delta_normalstate}
\end{align}

\section{\label{subsec:Stability}Phase diagram and stability of the model
in the scalar state}

The structure of the phase diagram in the scalar state, which is obtained
by solving the gap equations (\ref{eq:GE_m_normalstate}) and (\ref{eq:GE_Delta_normalstate}),
depends on the choice of $G$, $H$ and the cutoff $\Lambda$. In realistic NJL
parameter fitting, these model parameters are partially controlled
by fitting model predictions to lattice results, nuclear matter, and
meson properties at low baryon density. Since our model is purely
schematic and has reduced color and flavor degrees of freedom, we
base our choice of $G$ and $H$, in terms of $\Lambda$, on only two
requirements: (1) there emerges a relatively extensive coexistence
phase connecting the vacuum chiral symmetry breaking phase and high
density BCS phase, in order to mimic the realistic QCD phase diagram,
and (2) the system remains stable throughout the phase diagram. After
discussing the ranges of $G$ and $H$ consistent with these requirements,
we construct the phase diagram in the end of this section. 

We first address constraints on $G$ in the absence of pairing, i.e.,
$H=0$, $\Delta=0$. Then, for $M\neq0$ Eq.\,(\ref{eq:GE_m_normalstate})
becomes:
\begin{equation}
\frac{1}{2G}=\frac{1}{\pi^{2}}\int_{p_{F}}^{\Lambda}\frac{p^{2}dp}{\sqrt{M^{2}+p^{2}}}.\label{eq:GE_m_simple}
\end{equation}
The integral has an upper bound for all $M$. Therefore, at any given
Fermi momentum $p_{F}$, there is a minimum value for $G$, below
which the chiral condensate $\sim M$ cannot develop; the minimum
value can be evaluated by taking the limit $M\to0$ in Eq.\,(\ref{eq:GE_m_simple})
while regarding $G$ as a function of $p_{F}$:
\begin{equation}
G=\frac{\pi^{2}}{\Lambda^{2}-p_{F}^{2}}.\label{eq:Gc1}
\end{equation}
In particular, in the vacuum, $p_{F}=0$, one must have $G>\pi^{2}/\Lambda^{2}$
to have a non-vanishing $M$. We denote this lower bound as $G_{c1}=\pi^{2}/\Lambda^{2}$.

\begin{figure}
\includegraphics[scale=0.35]{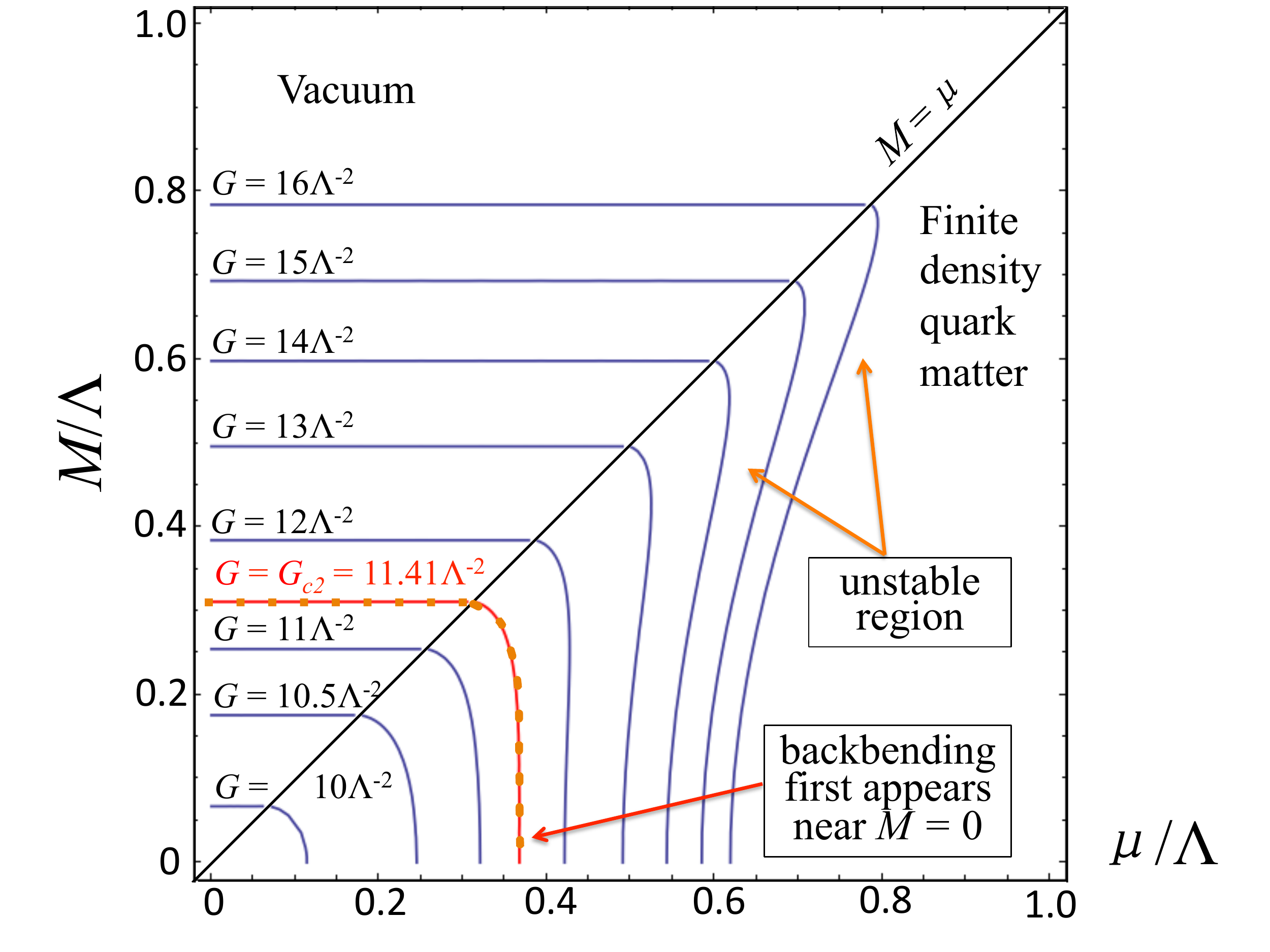}\caption{\label{fig:m_vs_mu_diffG}(Color online) Solutions to gap equation $M(\mu)$ for
varying $G$. Backbending indicating instability first occurs
at $G_{c2}$.}
\end{figure}

In addition the requirement of stability under density fluctuations
places an upper bound on $G$. Such stability requires $\partial\mu/\partial n>0$
where $n=p_{F}^{3}/3\pi^{2}$ is the quark density. This condition
can be related to the solution for $M(\mu)$ in Eq.\,(\ref{eq:GE_m_simple}).
Differentiating $p_{F}(\mu)^{2}=\mu^{2}-M(\mu)^{2}$ with respect to
$\mu$ we find
\begin{equation}
\frac{\partial n}{\partial\mu}=\frac{p_{F}}{\pi^{2}}\left(\mu-M\frac{\partial M}{\partial\mu}\right),\label{eq:nmu_stability}
\end{equation}
which must remain positive to ensure stability. From the plots of
the solutions $M(\mu)$ as a family of curves given for varying $G$
in Fig.\,\ref{fig:m_vs_mu_diffG}, we see that above a certain value
of $G$, the $M(\mu)$ curve begins to bend back\footnote{A similar instability related to back-bending of $\langle\bar{q}q\rangle(\mu)$ also appears in lattice gauge analyses of chiral restoration, e.g., \citep{Bilic1992}.}. When backbending
begins with increasing $G$, $\partial M/\partial\mu$, at first finite and negative, becomes
$-\infty$, turns to $+\infty$
and then becomes finite and positive. During backbending, Eq.\,(\ref{eq:nmu_stability})
cannot remain positive. As a result, the system becomes unstable against
density perturbations, and a homogeneous mean field solution for the
scalar state is unphysical.

\begin{figure}
\includegraphics[scale=0.35]{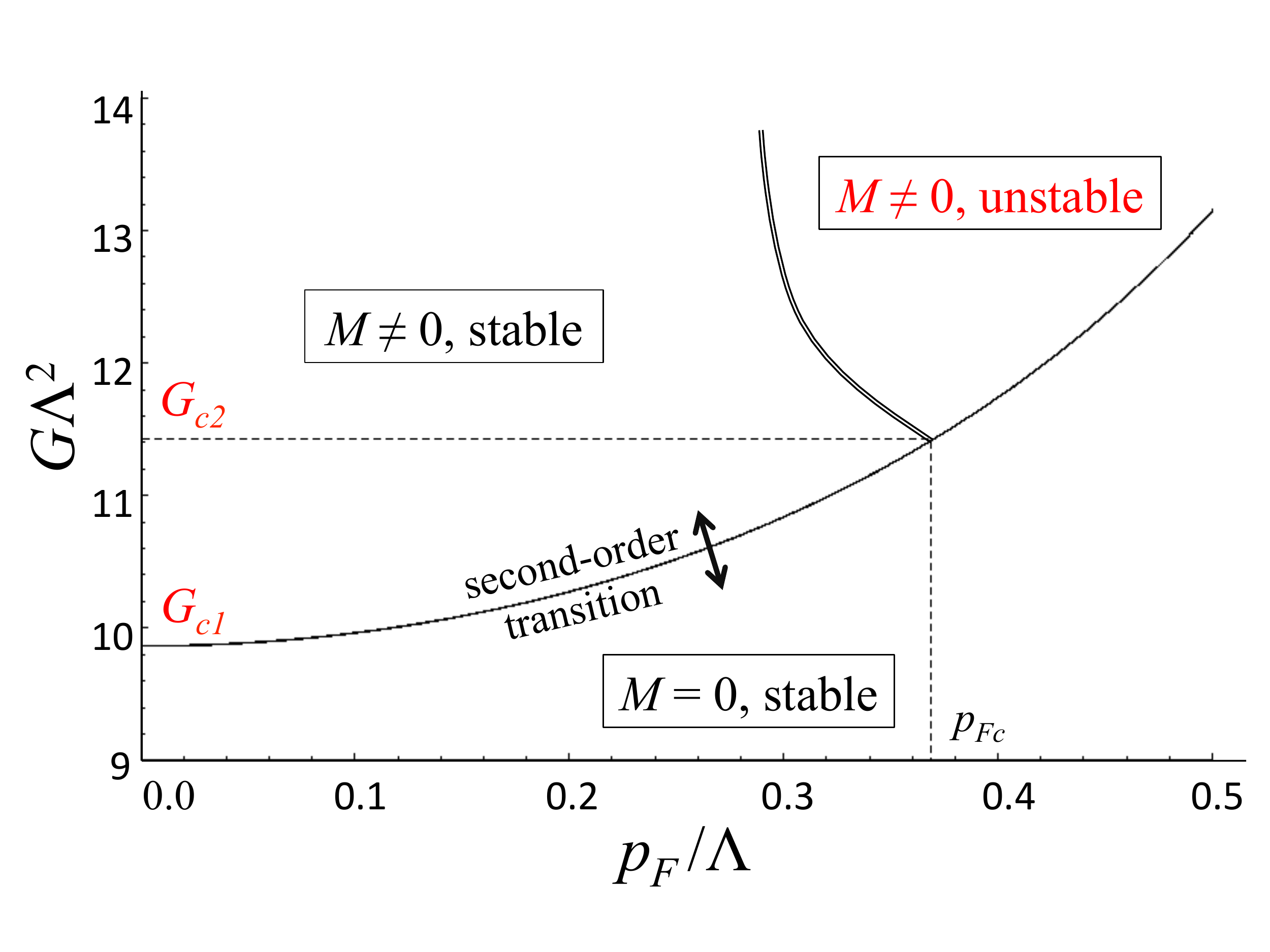}
\caption{\label{fig:stability}(Color online) Stability of the system at varying Fermi momentum
$p_{F}$ and $G$, in terms of $\Lambda$. In the range $G_{c1}<G<G_{c2}$,
the system is stable with a chirally broken vacuum.}
\end{figure}

To compute this upper bound $G_{c2}$ for $G$ above which backbending of
$M(\mu)$ happens, we observe that the stability
is first violated, with increasing $G$, for $M\to0$. In this limit,
$\partial M/\partial\mu$ can be calculated by differentiating Eq.\,(\ref{eq:GE_m_simple})
with respect to $\mu$, with the result
\begin{equation}
M\frac{\partial M}{\partial\mu}=\left(1-\ln\frac{\Lambda}{p_{F}(\mu)}\right)^{-1}\mu.
\end{equation}
The backbending-related divergence of $\partial M/\partial\mu$ then
appears at critical Fermi momentum $p_{Fc}$ obeying $1-\ln\Lambda/p_{Fc}=0$,
i.e., $p_{Fc}=\Lambda/e$ where $e$ is Napier's constant. Substituting
$p_{Fc}$ back into Eq.\,(\ref{eq:GE_m_simple}) together with $M\to0$,
we find the critical value,
\begin{equation}
G_{c2}=\frac{\pi^{2}}{(1-e^{-2})\Lambda^{2}}.
\end{equation}
Finally, we plot the stability of the system at varying Fermi momentum and 
$G$ in Fig.\,\ref{fig:stability}.
In the range $G_{c1}<G<G_{c2}$, the system always undergoes a smooth
second order transition from the chirally broken region $M\neq0$ to the
restored region $M=0$.

We now turn to constraints on $H$. Unlike in Eq.\,(\ref{eq:GE_m_normalstate})
for $M$, the integral in Eq.\,(\ref{eq:GE_Delta_normalstate}) for
$\Delta$ does not have an upper bound with varying $\Delta$, owing
to the singularity in $1/\omega_{+}(\bm{p})$ at the Fermi surface
$|\bm{p}|=p_{F}$ when $\Delta\to0$.  As a result, at any density
Eq.\,(\ref{eq:GE_Delta_normalstate}) always has a non-trivial solution
for all $H$, as in non-relativistic BCS theory.  Thus, diquark
pairing always appears at finite densities; neither a lower nor upper bound
for $H$ is imposed by requiring diquark pairing in the model.

The requirement of emergence of a coexistence phase, however, does
constrain $H$. For non-zero $M$ and $\Delta$, one can divide
the gap equation (\ref{eq:GE_m_normalstate}) by $M$ and (\ref{eq:GE_Delta_normalstate})
by $\Delta$, and subtract one from the other, to find
\begin{equation}
\frac{1}{2H}-\frac{1}{2G}=\int_{\bm{p}}\frac{\mu}{\sqrt{\bm{p}^{2}+M^{2}}}\left(\frac{1}{\omega_{+}(\bm{p})}-\frac{1}{\omega_{-}(\bm{p})}\right).\label{eq:stability_H-G}
\end{equation}
The right side of Eq.\,(\ref{eq:stability_H-G}) is always positive
since $\omega_{+}(\bm{p})<\omega_{-}(\bm{p})$.   As a consequence,
one must have $H<G$ to have a coexistence region.

\begin{figure}
\includegraphics[scale=0.35]{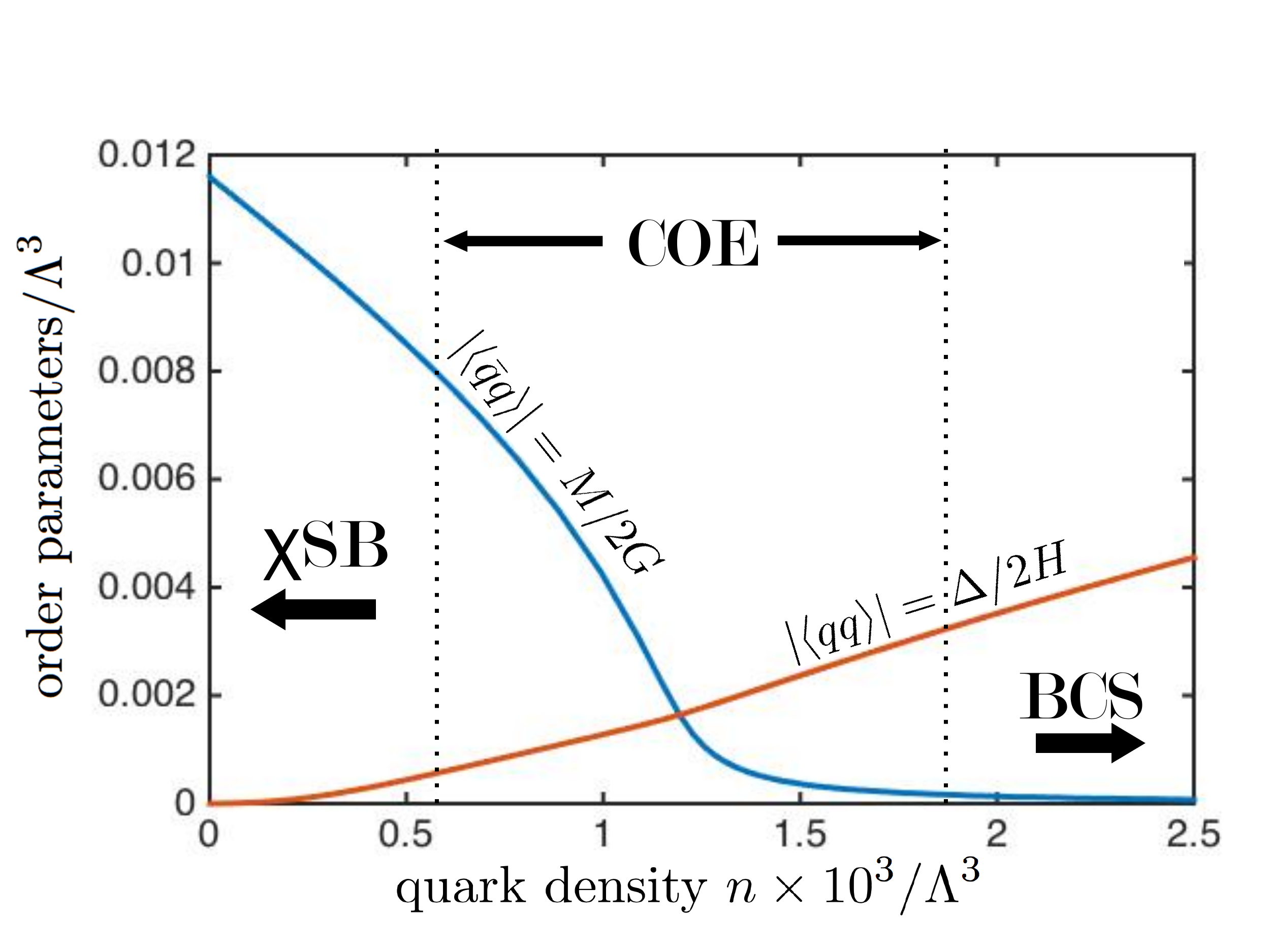}
\caption{\label{fig:phase}(Color online) The evolution of $|\langle\bar{q}q\rangle|=M/2G$
and $|\langle qq\rangle|=\Delta/2H$ against quark density $n$ with
$G=11\Lambda^{-2}$ and $H=6\Lambda^{-2}$.  The phase diagram can be roughly divided
into the chirally broken vacuum ($\chi$SB) with $\Delta\approx0$,
$M\protect\neq0$, the coexistence (COE) phase where $M$ and $\Delta$
are both finite and comparable, and the high density BCS limit where
$\Delta\protect\neq0$ but $M\approx0$.}
\end{figure}

For $H\neq0$ and $G=0$, the system is always stable as in non-relativistic
BCS. For finite $G$ however, proving stability becomes subtle and
unfortunately too algebraically overwhelming to analyze by hand. Numerical
calculation suggests that instability could still develop when $H$
becomes comparable to $G$, but for relatively small $H$, $\lesssim G/2$,
a stable coexistence region can be achieved. Figure \ref{fig:phase}
shows the phase structure of the model at varying $p_{F}$ plotted
for a good choice $G=11\Lambda^{-2}$ and $H=6\Lambda^{-2}$.

In the following we discuss the collective modes of the system assuming
a phase structure as in Fig.\,\ref{fig:phase}.

\section{\label{subsec:Collective_modes}Collective modes: mass spectra and
decay constants}

In this section we identify the collective excitations present in
the model system.   The following discussion is valid for general phase between the scalar and pseudoscalar condensates, not just for the scalar state.  In the chiral limit, the collective excitations include two NG modes associated with the spontaneous breakings
of the $U(1)_{L}\otimes U(1)_{R} = U(1)_V\otimes U(1)_A$ symmetries -- the pionic mode $\pi_G$, which is a linear combination of the vacuum pion mode $\pi$ and
the diquark-condensate pion mode $\tilde \pi$, and a phonon
mode corresponding to fluctuations of the overall phase of the scalar and pseudoscalar pairing gaps, $\Delta_s$ and $\Delta_{ps}$.  
In addition the system has 
four massive modes, one corresponding to the other linearly independent mixture of $\pi$ and $\tilde\pi$,  two Higgs-like modes corresponding to the fluctuations of the magnitudes of the chiral and diquark condensates, and finally one corresponding to the relative phase of the scalar and pseudoscalar condensates.  The modes are summarized in Table \ref{tb:modes}. 

\begin{center}
\begin{table}
\begin{tabular}{|c|l|c|c|}
\hline 
Mode & Description & Parity\tabularnewline
\hline 
\hline 
$\theta_{B}$ & phonon; NG boson of broken $U(1)_{V}$  & +\tabularnewline
\hline 
$\theta_{\pi}+\theta_{d}$ & pionic mode; NG boson of broken $U(1)_{A}$  & -\tabularnewline
\hline 
$\theta_{\pi}-\theta_{d}$ & massive chiral oscillation between $\pi$ and $\Delta_{ps}$ & -\tabularnewline
\hline 
$M$ & Higgs-like; breaks $U(1)_{A}$ & +\tabularnewline
\hline 
$\Delta$ & Higgs-like; breaks $U(1)_{V}$ and $U(1)_{A}$ & +\tabularnewline
\hline 
$\phi$ & relative phase oscillation between $\Delta_{s}$ and $\Delta_{ps}$ & +\tabularnewline
\hline
\end{tabular}
\caption{Six normal collective modes of the system.} 
\label{tb:modes}
\end{table}
\end{center}

\subsection{\label{subsec:mode_para}General parametrization of the collective modes}

The collective modes in the model system can be directly obtained
via the variation of $\Omega$ under small fluctuations of the mean
fields. To parametrize the modes, we write the mean fields in terms
of the chiral sector axial $U(1)_{A}$ angle $\theta_{\pi}$, the diquark sector $U(1)_{A}$
angle $\theta_{d}$, the relative phase angle $\phi$ between $\Delta_{s}$
and $\Delta_{ps}$, and the overall $U(1)_{V}$ phase angle $\theta_{B}$:
\begin{eqnarray}
\sigma & = & -M\cos\theta_{\pi},\nonumber \\
\pi & = & -M\sin\theta_{\pi},\nonumber \\
\Delta_{s} & = & -i\Delta e^{i\theta_{B}}e^{i\phi/2}\cos\theta_{d},\nonumber \\
\Delta_{ps} & = & i\Delta e^{i\theta_{B}}e^{-i\phi/2}\sin\theta_{d}.\label{eq:MF_para}
\end{eqnarray}
The oscillations of $\theta_\pi$ correspond to the usual pion mode, $\pi$, while those of $\theta_d$ correspond
to the diquark-condensate pion, $\tilde\pi$.
We choose  $\Delta>0$,
 $M>0$, and thus $\sigma<0$ at $\theta_{\pi}=0$ (see Sec.\,\ref{subsec:mode_mq}).
The $U(1)_{V}$ transformation is trivial, with both $\Delta_{s}$
and $\Delta_{ps}$ picking up the same phase $\theta_{B}\rightarrow\theta_{B}+\theta_{V}$.
On the other hand, when rotating the system by a $U(1)_{A}$ angle
$\theta_{A}$, the $\sigma$ and $\pi$ fields transform as
\begin{eqnarray}
\left.\begin{array}{c}
\sigma\rightarrow\sigma\cos\theta_{A}-\pi\sin\theta_{A}\\
\pi\rightarrow\pi\cos\theta_{A}+\sigma\sin\theta_{A}
\end{array}\right\}  & \Rightarrow & \theta_{\pi}\rightarrow\theta_{\pi}+\theta_{A}.\label{eq:thetapi_trans}
\end{eqnarray}

    However the transformation of $\Delta_{s}$ and $\Delta_{ps}$  is
more complicated:
\begin{eqnarray}
\Delta_{s} & \rightarrow & \Delta_{s}\cos\theta_{A}+\Delta_{ps}\sin\theta_{A}\nonumber \\
 & = & -i\Delta\left[\cos\frac{\phi}{2}\cos\left(\theta_{d}+\theta_{A}\right)+i\sin\frac{\phi}{2}\sin\left(\theta_{d}-\theta_{A}\right)\right]\nonumber \\
 & \equiv & -i\Delta e^{i\phi'/2}\cos\theta_{d}',\nonumber \\
\Delta_{ps} & \rightarrow & \Delta_{ps}\cos\theta_{A}-\Delta_{s}\sin\theta_{A}\nonumber \\
 & = & -i\Delta\left[-\cos\frac{\phi}{2}\sin\left(\theta_{d}+\theta_{A}\right)+i\sin\frac{\phi}{2}\sin\left(\theta_{d}-\theta_{A}\right)\right]\nonumber \\
 & = & i\Delta e^{-i\phi'/2}\sin\theta_{d}',
 \label{eq:Delta_trans}
\end{eqnarray}
that is, both the relative phase $\phi$ and the chiral angle $\theta_{d}$
change under the chiral transformation. When the
two condensates are in phase, i.e., $\phi=0$, the result reduces to $\theta_{d}\rightarrow\theta_{d}+\theta_{A}$
and  $\theta_{\pi} \rightarrow\theta_{\pi}+\theta_{A}$.  In this case, the diquark-condensate pion corresponds to oscillations of
the product of the two diquark terms, $\Delta_s\Delta^*_{ps}$.   For non-zero $\phi$, we have 
\begin{eqnarray}
\cos\theta_{d} & \rightarrow & \cos\theta_{d}'=\left[\cos^{2}\frac{\phi}{2}\cos^{2}\left(\theta_{d}+\theta_{A}\right)\right.\nonumber \\
 &  & \hspace{1.5cm}\left.+\sin^{2}\frac{\phi}{2}\sin^{2}\left(\theta_{d}-\theta_{A}\right)\right]^{\frac{1}{2}},\nonumber \\
\phi & \rightarrow & \phi'=2\tan^{-1}\left[\frac{\tan(\phi/2)\sin\left(\theta_{d}-\theta_{A}\right)}{\cos\left(\theta_{d}+\theta_{A}\right)}\right].\label{eq:Delta_paratrans}
\end{eqnarray}
In terms of the parametrization (\ref{eq:MF_para}), the four invariants (\ref{eq: 4inv}) become
\begin{align}
 & \sigma^{2}+\pi^{2}=M^{2},\nonumber \\
 & |\Delta_{s}|^{2}+|\Delta_{ps}|^{2}=\Delta^{2},\nonumber \\
 & \mbox{Im}\left[\Delta_{s}\Delta_{ps}^{*}\right]=\frac{\Delta^{2}}{2}\sin2\theta_{d}\sin\phi,\label{eq:3invariants}
\end{align}
and
\begin{align}
 & |\sigma\Delta_{ps}+\pi\Delta_{s}|^{2}\nonumber \\
 & =M^{2}\Delta^{2}\left[\sin^{2}\left(\theta_{\pi}-\theta_{d}\right)-2\sin\theta_{\pi}\sin\theta_{d}\sin^{2}\frac{\phi}{2}\right].\label{eq:1invariant}
\end{align}
From the six independent real degrees of freedom, $M,\Delta,\theta_{\pi},\theta_{d},\theta_{B},\phi$, we
identify the six independent normal modes:

1) The massless phonon mode, corresponding to fluctuations of $\theta_B$.  This mode is massless since the free energy does not depend on this angle.  

2) The massless pionic mode, $\pi_G$, identified with fluctuations of the angle $\theta_G \equiv (\theta_\pi + \theta_d)/2$.  Again the free energy does not depend on $\theta_G$.   This 
mode describes the simultaneous chiral rotation of $\sigma$ and $\Delta_{s}$
in the same direction and is the NG mode.

3) A massive pionic mode, denoted as $\pi_M$, identified with fluctuations of $\theta_M \equiv (\theta_\pi - \theta_d)/2$. This mode does not correspond to a $U(1)_{A}$ rotation of the system
and is thus always massive.  The stiffness term for this mode is
\begin{align}
\frac{\partial\Omega}{\partial\sin^{2}\left(\theta_{\pi}-\theta_{d}\right)} & =-\sum_{\pm}\int_{\bm{p}}\frac{M^{2}\Delta^{2}}{\pm\delta(\bm{p})\omega_{\pm}}\label{eq:piM_mass}\\
 & =\int_{\bm{p}}\frac{M^{2}\Delta^{2}}{\mu\sqrt{\bm{p}^{2}+M^{2}}}\left(\frac{1}{\omega_{+}}-\frac{1}{\omega_{-}}\right),\nonumber 
\end{align}
which is always positive in the coexistence phase, where $M^{2}\Delta^{2}\neq0$.  The squared mass of the massive mode, proportional to the stiffness term, $\sim M^{2}\Delta^{2}$, indicates that the mixing naturally
occurs as long as there is a coexistence phase, even without any explicit
$\bar{q}q$-$qq$ coupling interactions at mean field level in the
Lagrangian.
This massive excitation always accompanies the chiral NG mode $\pi_G$; however it becomes unstable against decay into $\pi_G$ when higher order fluctuations of the fields are taken into account.    The mixing of the $\pi$ and $\tilde\pi$ modes to form  the massless $\pi_G$ and the massive $\pi_M$ modes
is a concrete example of the mechanism described in \citep{hatsudaCollective}.  Here mixing results from the term $|\sigma\Delta_{ps}+\pi\Delta_{s}|^{2}$, which leads to terms $\sim\pi\tilde{\pi}$.

4) The two massive modes corresponding to fluctuations of $\Delta$ and $M$.   These modes can be associated with oscillations in the radial direction of `Mexican hat' potentials describing
the broken symmetry state.   In particular, the fluctuations of $M$
are related to the heavy $\sigma$-meson in nuclear matter.

5) The massive mode associated with fluctuations of the relative phase $\phi$.
This mode is generally
not discussed in NJL investigations of the phase diagram.  If one starts
in the scalar state with $\phi=0$, a axial rotation $\theta_{A}$
will leave this angle untouched, as seen from Eq.\,(\ref{eq:Delta_paratrans}).
Note that if either $\Delta_{s}$ or $\Delta_{ps}$ vanishes,
this mode is not present.

   Having delineated the modes, we study in detail the transition from the vacuum pion mode $\pi$ associated with $\theta_{\pi}$ to the diquark
pion mode $\pid$ in BCS phase at high density associated with $\theta_{d}$.  We consider the fluctuations of the system about the scalar
state fixing $\phi=0$, and neglect the phonon mode as well as the massive modes $M$ and $\Delta$; the latter of positive parity do not mix with the pionic modes.

\subsection{\label{subsec:mode_mass}The mass matrix for $\pi$ and $\tilde\pi$}

We first calculate the two-by-two mass matrix $\M$ relating the $\pi$ and $\tilde{\pi}$ modes in an effective Lagrangian.  To do so we expand the free energy $\Omega$ in terms of $\theta_\pi$ and $\theta_d$ to second order.  As discussed earlier, of the two new linearly independent modes, $\pi_G$ and $\pi_M$, the NG mode $\pi_G$ remains massless while $\pi_M$ must be massive.  In fact, Eq.~(\ref{eq:piM_mass})  shows that
\begin{align}
\Omega & =\Omega_{0}+\frac{1}{2}\theta_{M}^{2}\int_{\bm{p}}\sum_{\pm}\frac{1}{\omega_{\pm}(\bm{p})\epsilon_{\pm}(\bm{p})\mu}+\ldots \nonumber \\
 & \equiv\Omega_{0}+\frac{1}{2}\vec{\theta}\,^{T} \Xi  \vec{\theta}+\ldots
 \label{eq:Mez}
\end{align} 
where the vector $\vec{\theta} \equiv (\theta_\pi, \theta_d)^T$, and $\Omega_0 = \Omega(\theta_\pi=\theta_d=0)$.  Equation (\ref{eq:Mez}) immediately indicates that the stiffness matrix for the angles $\vec{\theta}$ is
\begin{equation}
\Xi
=M^{2}\Delta^{2}a\left(\begin{array}{cc}
1 & -1\\
-1 & 1
\end{array}\right);
\label{eq:M_res}
\end{equation}
where 
\begin{equation}
a(\mu)\equiv \int_{\bm{p}}\sum_{\pm}\frac{1}{\omega_{\pm}(\bm{p})\epsilon_{\pm}(\bm{p})\mu},
\label{eq:a}
\end{equation}
which is always positive.  The matrix $\Xi$ is related to the mass matrix $\M$ for the two pionic fields, i.e., the vacuum pion $\pi = f_\pi \theta_\pi$ and the diquark-condensate pion $\tilde{\pi} = f_{\tilde{\pi}} \theta_d$ ($f_\pi$ and $f_{\tilde{\pi}}$ being their decay constants), by
\begin{align}
\M=\mathcal{F}^{-1}\Xi\mathcal{F}^{-1}=M^{2}\Delta^{2}a\left(\begin{array}{cc}
1/f_{\pi}^{2} & -1/f_{\pi}f_{\tilde{\pi}}\\
-1/f_{\pi}f_{\tilde{\pi}} & 1/f_{\tilde{\pi}}^{2}
\end{array}\right),
\label{eq:M_GB}
\end{align}
where $\mathcal{F} = \mbox{diag}(f_\pi,\, f_{\tilde{\pi}})$ is a simple invertible matrix relating $\pi$ and $\tilde{\pi}$ to $\theta_\pi$ and $\theta_d$:
\begin{align}
\vec{\pi}\equiv\left(\begin{array}{c}
\pi\\
\tilde{\pi}
\end{array}\right)=\mathcal{F}\left(\begin{array}{c}
\theta_{\pi}\\
\theta_{d}
\end{array}\right).
\label{eq:pi_to_theta}
\end{align}

The mass matrix $\M$ in Eq.~(\ref{eq:M_GB}) is diagonalized by the transformation
\begin{equation}
\left(\begin{array}{c}
\pi\\
\tilde{\pi}
\end{array}\right)=\frac{1}{\sqrt{f_{\pi}^{2}+{f}_{\tilde\pi}^{2}}}\left(\begin{array}{cc}
f_{\pi} & {f}_{\tilde\pi}\\
{f}_{\tilde\pi} & -f_{\pi}
\end{array}\right)\left(\begin{array}{c}
\pi_G\\
\pi_M
\end{array}\right),\label{eq:piGBtopi_1}
\end{equation}
directly relating the $\pi_G$ and $\pi_M$ modes to the initial $\pi$ and $\pid$ modes, with the expected mixing ratio described in Ref.~\citep{hatsudaCollective}.
The two eigenvalues of $\M$, $m_G^2 = 0$ and $m_{M}^{2}=M^{2}\Delta^{2}a \left(f_{\pi}^{-2}+f_{\tilde{\pi}}^{-2}\right)$,
give the masses of $\pi_G$ and $\pi_M$.

    The off-diagonal terms in $\M$, corresponding to mixing of  the $\pi$ and $\tilde{\pi}$ modes, can also be understood in terms of perturbing the correlation functions in the chiral and diquark channel.  Essentially, the corresponding off-diagonal term in $\Xi$ can be written as
\begin{align}
\Xi_{12} & =\frac{\partial}{\partial\theta_{d}}\left(\frac{\partial\Omega}{\partial\theta_{\pi}}\right)=-iM\frac{\partial\langle\bar{\psi}\Gamma_{\pi}\psi\rangle}{\partial\theta_{d}} \nonumber \\
 & =M\Delta\frac{\partial\langle\bar{\psi}\Gamma_{\pi}\psi\rangle}{\partial\langle\bar{\psi}\Gamma_{\Delta_{ps}}\psi\rangle}\sim\frac{\partial\langle\bar{q}i\gamma_{5}q\rangle}{\partial\langle qq\rangle}\Big|_{\theta_\pi},
\end{align}
where we have defined the matrices in Nambu-Gor'kov-Dirac space
\begin{align}
\Gamma_{\pi} & \equiv\frac{1}{M}\frac{\partial S^{-1}}{\partial\theta_{\pi}}=\left(\begin{array}{cc}
i\gamma_{5} & 0\\
0 & i\gamma_{5}
\end{array}\right), \nonumber \\
 \Gamma_{\Delta_{ps}} & \equiv \frac{1}{i\Delta}\frac{\partial S^{-1}}{\partial\theta_{d}}=\left(\begin{array}{cc}
0 & 0\\
\mathbbm{1}& 0
\end{array}\right),
\label{eq:GammaPi}
\end{align}
with $\mathbbm{1}$ the ${4\times4}$ identity matrix in Dirac space.

   It is instructive to compare the results (\ref{eq:M_res}) and (\ref{eq:M_GB}) with the general discussion in \citep{hatsudaCollective}, where the pion mass matrix for $\pi$ and $\tilde{\pi}$ was constructed from a general Ginzburg-Landau expansion of the free energy based on symmetry principles.  Up to fourth order terms in the mean fields, we see that the existence of the mixing terms, i.e., the off-diagonal terms in $\M$, $\sim M^{2}\Delta^{2}$ falls naturally out of the present expansion of the free energy.  The existence of these terms can be understood as a consequence of Goldstone's theorem, since only the NG boson mode should remain massless; individual fluctuations of $\theta_\pi$ and $\theta_d$ no longer correspond to a global $U(1)_A$ transformation of the system, thus they cannot remain massless in the coexistence phase. Only the re-diagonalized mode $\pi_G$ corresponding to the simultaneous rotation of $\theta_\pi$ and $\theta_d$ is massless, i.e., the mode $\theta_G = (\theta_\pi + \theta_d)/2$. 

\subsection{\label{subsec:mode_decay}The decay constant of the chiral NG mode $\pi_G$}

  Having identified the mass matrix, we next study the decay constant of the NG mode $\pi_G$, which can be identified as the kinetic energy coefficient of $\theta_G$ in the effective Lagrangian of the bosonic fields in the long wavelength limit.  To do so, we consider spatially dependent fluctuations of $\theta_\pi$ and $\theta_d$.
We first apply a Hubbard-Stratonovich transformation of the original quark system
into a coupled system of quark fields and bosonic 
fields corresponding to the fluctuations of all the mean fields $\sigma$, $\pi$, $\Delta_{s}$,
and $\Delta_{ps}$. We denote these fluctuations by $\hat{\sigma}$, $\hat{\pi}$, $\hat{\Delta}_{s}$,
and $\hat{\Delta}_{ps}$, where the hat distinguishes the bosonic field fluctuations from their corresponding mean field values.

  The partition function can be computed from the functional path integral:
\begin{align}
\mathcal{Z} & =\int dq\,d\bar{q}\,d\hat{\sigma}\,d\hat{\pi}\,d\hat{\Delta}_{s}\,d\hat{\Delta}_{s}^{*}\,d\hat{\Delta}_{ps}\,d\hat{\Delta}_{ps}^{*}\nonumber \\
 & \ \ \times\exp\left\{ i\int d^{4}x\left[\bar{q}S^{-1}q-V(\hat{\sigma},\hat{\pi},\hat{\Delta}_{s},\hat{\Delta}_{ps})\right]\right\} ,\label{eq:Z}
\end{align}
where $t = i\tau$, with $0\le\tau\le \beta$, $\beta$ being the inverse temperature.   The inverse quark propagator $S^{-1}$ is perturbed from that in the scalar state
 $S_{0}^{-1}$, defined in Eq.\,(\ref{eq:InvProp_normalstate}),
by the bosonic fields:
\begin{align}
S^{-1} & =S_{0}^{-1}+\hat{x},
\label{eq:Sinv_pert}
\end{align}
where
\begin{align}
\hat{x} & =\left(\begin{array}{cc}
\hat{\sigma}+i\gamma_{5}\hat{\pi} & i\gamma_{5}\hat{\Delta}_{s}^{*}+\hat{\Delta}_{ps}^{*}\\
i\gamma_{5}\hat{\Delta}_{s}+\hat{\Delta}_{ps} & \hat{\sigma}+i\gamma_{5}\hat{\pi}
\end{array}\right).\label{eq:xdef}
\end{align}
The potential term is 
\begin{align}
V(\hat{\sigma},\hat{\pi},\hat{\Delta}_{s},\hat{\Delta}_{ps}) & =\frac{1}{4G}\left[\left(\hat{\sigma}-M\right)^{2}+\hat{\pi}^{2}\right]\nonumber \\
 & +\frac{1}{4H}\left[|\hat{\Delta}_{s}-i\Delta|^{2}+|\hat{\Delta}_{p}|^{2}\right].\label{eq:VinZ}
\end{align}

   Integrating out the quark fields to obtain the determinant of $S^{-1}$,
and then re-exponentiating we find the effective action $\mathcal{A}$
involving only the bosonic fields:  
\begin{align}
\mathcal{A} & =-i\mbox{Tr}\ln S^{-1}+\int d^{4}x\,V \nonumber \\
 & =-i\mbox{Tr}\ln S_{0}^{-1}-i\mbox{Tr}\left[S_{0}\hat{x}-\frac{(S_{0}\hat{x})^2}{2}\right]+\int d^{4}x\,V+\ldots, \label{eq:A_expansion}
\end{align}
where ``$\mbox{Tr}$'' denotes the sum over
all indices, including position (or equivalently, momentum).   In the following we drop the constant term $-i\mbox{Tr}\ln S_{0}^{-1}$ 
as it does not involve the bosonic fluctuations.  We write the bosonic field fluctuations in
terms of the spatially dependent real bosonic fields $\hat{\theta}_{\pi}$ and $\hat{\theta}_d$,  as in Eq.~(\ref{eq:MF_para}): 
\begin{align}
  -M \cos \hat{\theta}_\pi & =  \hat{\sigma} - M, \nonumber \\
 - M \sin \hat{\theta}_\pi & =  \hat{\pi} , \nonumber \\
 - i\Delta \cos \hat{\theta}_d & =  \hat{\Delta}_s - i\Delta , \nonumber \\
  i\Delta \sin \hat{\theta}_d & =  \hat{\Delta}_{ps}.
\label{eq:theta_hatdef} 
\end{align}
As a result, to leading order in $\hat{\theta}_{\pi}$ and $\hat{\theta}_d$, 
\begin{equation}
\hat{\sigma}\approx\frac{1}{2}M\hat{\theta}_{\pi}^{2},\,\hat{\pi}\approx-M\hat{\theta}_{\pi},\,\hat{\Delta}_{s}\approx\frac{i}{2}\Delta\hat{\theta}_{d}^{2},\,\hat{\Delta}_{ps}\approx i\Delta\hat{\theta}_{d};\label{eq:MFtotheta}
\end{equation}
using this equation we expand $\mathcal{A}$ up to second order in
$\hat{\theta}_{\pi}$ and $\hat{\theta}_{d}$, writing first,  
\begin{align}
\hat{x}  \approx & \, M\left(\begin{array}{cc}
\frac{1}{2}\hat{\theta}_{\pi}^{2}-i\gamma_{5}\hat{\theta}_{\pi} & 0\\
0 & \frac{1}{2}\hat{\theta}_{\pi}^{2}-i\gamma_{5}\hat{\theta}_{\pi}
\end{array}\right)\nonumber \\
&  +\Delta\left(\begin{array}{cc}
0 & \frac{1}{2}\gamma_{5}\hat{\theta}_{d}^{2}-i\hat{\theta}_{d}\\
-\frac{1}{2}\gamma_{5}\hat{\theta}_{d}^{2}+i\hat{\theta}_{d} & 0
\end{array}\right)\nonumber \\
 \equiv & \, M\left(\frac{1}{2}\hat{\theta}_{\pi}^{2}\Gamma_{\sigma}-\hat{\theta}_{\pi}\Gamma_{\pi}\right)+\Delta\left(\frac{1}{2}\hat{\theta}_{d}^{2}\Gamma_{\sigmad}-\hat{\theta}_{d}\Gamma_{\pid}\right),\label{eq:xtotheta}
\end{align}
where the matrices $\Gamma_\sigma$, $\Gamma_{\sigmad}$, and $\Gamma_{\pid}$  in Nambu-Gor'kov-Dirac space are
\begin{align}
\Gamma_{\sigma} & =\left(\begin{array}{cc}
\mathbbm{1} & 0\\
0 & \mathbbm{1}
\end{array}\right),
\Gamma_{\sigmad}  =\left(\begin{array}{cc}
0 & \gamma_{5}\\
-\gamma_{5} & 0
\end{array}\right), \nonumber \\
\Gamma_{\pid}&=\left(\begin{array}{cc}
0 & i\mathbbm{1}\\
-i\mathbbm{1} & 0
\end{array}\right),
\label{eq:Vertix}
\end{align}
while $\Gamma_\pi$ is already defined in Eq.~(\ref{eq:GammaPi}).

  In terms of real vector field $\vec{\theta}\equiv\left(\hat{\theta}_{\pi},\hat{\theta}_{d}\right)^{T}$, the quadratic effective action becomes
\begin{equation}
\mathcal{A}\approx\frac{1}{2}\beta \mathcal{V}\int\frac{d^{4}k}{(2\pi)^{4}}\,{\vec{\theta}}(-k)^T\mathcal{D}_\theta^{-1}(k){\vec{\theta}}(k),\label{eq:A_qua}
\end{equation}
where
\begin{align}
&\mathcal{D}_\theta^{-1}(k) \nonumber\\ &\hspace{6pt}=\left(\begin{array}{cc}
M^{2}\left(B_{\pi\pi}(k)-{1/2G}\right) & M\Delta B_{\pi d}(k)\\
M\Delta B_{\pi d}(k) & \Delta^{2}\left(B_{dd}(k)-{1/2H}\right)
\end{array}\right)
\label{eq:Pi}
\end{align}
is a two-by-two matrix, and $\mathcal{V}$ is the spatial volume of the system.
The bubbles are defined by\footnote{Note that with the $u$, $d$ quarks replaced by protons and neutrons the bubble $B_{\pi\pi}$ is simply the self-energy of the conventional in-nuclear medium pion Green's function.}
\begin{align}
B_{\pi\pi}(k) & =i\int\frac{d^{4}p}{(2\pi)^{4}}\mbox{tr}\left(S_{0}(p)\Gamma_{\pi}S_{0}(p-k)\Gamma_{\pi}\right),\nonumber \\
B_{\pi d}(k) & =i\int\frac{d^{4}p}{(2\pi)^{4}}\mbox{tr}\left(S_{0}(p)\Gamma_{\pi}S_{0}(p-k)\Gamma_{\pid}\right),\nonumber \\
B_{dd}(k) & =i\int\frac{d^{4}p}{(2\pi)^{4}}\mbox{tr}\left(S_{0}(p)\Gamma_{\pid}S_{0}(p-k)\Gamma_{\pid}\right),
\label{eq:bubbles}
\end{align}
where ``{\rm tr}'' denotes the Dirac and Nambu-Gor'kov trace.  The factors $1/2G$ and $1/2H$ result from using the gap equations (\ref{eq:GE_m_normalstate}) and (\ref{eq:GE_Delta_normalstate}).  Note that by the definition (\ref{eq:A_qua}), $\mathcal{D}_\theta^{-1}(0)$ simply reduces to $-\Xi$ at zero momentum $k=0$. 

At finite density, ${\cal D}_\theta^{-1}$ is generally not a function of the Lorentz scalar $k^2$.\footnote{Even in the vacuum use of a three-momentum cutoff violates Lorentz invariance.}  Thus
the temporal and spatial decay constants need not be equal at finite density.  To second order in $k$,
\begin{equation}
\mathcal{D}_\theta^{-1}(k)\approx- \Xi+{\mathcal{Q}}k_0^2 - \mathcal{Q}_v \bm{k}^2,
\end{equation}
where $\mathcal{Q}$ and $\mathcal{Q}_v$ are also two-by-two matrices. The dispersion relations of the modes are given by the
eigenvalues of $\mathcal{D}_\theta^{-1}$, the decay constants are contained
in the matrix ${\mathcal{Q}}$, and the mode velocities are included in ${\mathcal{Q}}_v$.  As we show shortly, after keeping only the leading order logarithm divergencies, $\mathcal{Q}$ is related to the matrix $\mathcal{F}$ as defined in Eq.~(\ref{eq:pi_to_theta}) by $\mathcal{Q}=\mathcal{F}^2$, while ${\mathcal{Q}}_v = \mbox{diag}(v_\pi^2, v_{\tilde{\pi}}^2)\mathcal{Q}$ (see Eq.~(\ref{eq:fpitoF})) where $v_\pi$ and $v_{\tilde{\pi}}$ are the mode velocities of $\pi$ and $\tilde{\pi}$. 

    The bubbles (\ref{eq:bubbles})
can be directly calculated from  Eq.\,(\ref{eq:Prop_normalstate}). To calculate the decay constant matrix $\mathcal{Q}$, we choose
$k=(k_{0},\bm{0})$, and then take derivatives of the bubbles with regard to $k_0$.
The $p_0$ integrals are Matsubara frequency summations with $p_0\to i\omega_\nu = 2\pi i T\nu$ and $\int dp_0 \to 2\pi  iT \sum_\nu$, where $\nu = \pm 1/2, \pm 3/2, \ldots$.
 In terms of the quasiparticle spectrum $\omega_\pm$, the free particle dispersion $\epsilon_\pm$, and the coherence functions $v_\pm$ and $u_\pm$ defined in Sec.~\ref{sec:scalar} -- all functions of the three-momentum integration variable $\bm p$ -- the bubbles are: 
 \begin{align}
B_{\pi\pi}(k_{0}^{2}) = & \int_{\bm{p}}\sum_{j,\ell=\pm}\left(u_{j}v_{\ell}-v_{j}u_{\ell}\right)^{2}\left(1-\frac{M^{2}+\bm{p}^{2}}{\epsilon_{j}\epsilon_{\ell}}\right)A_{j\ell}(k_{0}),\nonumber \\
B_{dd}(k_{0}^{2}) = & \int_{\bm{p}}\sum_{j,\ell=\pm}\left(v_{j}v_{\ell}+u_{j}u_{\ell}\right)^{2}\left(1-\frac{M^{2}-\bm{p}^{2}}{\epsilon_{j}\epsilon_{\ell}}\right)A_{j\ell}(k_{0}),\nonumber \\
B_{\pi d}(k_{0}^{2})  = & \int_{\bm{p}}\sum_{j,\ell=\pm}(v_{j}v_{\ell}+u_{j}u_{\ell})(v_{j}u_{\ell}-u_{j}v_{\ell}) \nonumber \\
& \times \frac{M(\epsilon_{\ell}-\epsilon_{j})}{\epsilon_{\ell}\epsilon_{j}}A_{j\ell}(k_{0}),
\label{eq:bubble_normalstate}
\end{align}
where 
\begin{align}
  A_{j\ell}(k_0)   =\frac{1}{2} \left( - \frac1{k_0 - \omega_{j} - \omega_{\ell}  } + \frac1{k_0+\omega_{j}+\omega_{\ell}} \right).
  \label{eq:Adef}
\end{align}

The physical interpretation of Eqs.~(\ref{eq:bubble_normalstate}) for the bubbles is the following.    The first factor, sums of products between coherence functions, indicates whether the quark loop connects the quark field with the quark field or with the charge-conjugate quark field.  The second factor, involving $\epsilon$'s, $M^2$ and $\bm{p}^2$, depends on whether the quark loop connects particle-antihole states with particle-antihole states, or with antiparticle-hole states.  Both the first and second factors are at most of order unity.  The final factor $A_{j\ell}$, Eq.~(\ref{eq:Adef}), reveals the pole structure of the external frequency $k_0$; it contains a pair of poles located at $\pm (\omega_j + \omega_\ell)$ with opposite signs for the corresponding residues, representing the pion state and anti-pion state described by the bubble.  In our model the pion is only neutral, thus they represent the same pion state.  In $N_f = 2$ models where charged pions are present, the dual poles would represent the $\pi^+$ and the $\pi^-$ state separably. 

  For example, consider the $B_{\pi\pi}$ bubble; the factor $(u_j v_\ell - v_j u_\ell)^2$ involves products between $v$ and $u$, indicating that the quark loop connects the quark field with the charge-conjugate field; the factor $1-(M^2+\bm{p}^2)/\epsilon_j\epsilon_\ell$ vanishes unless $j=-\ell$, indicating that the particle-antihole state is connected to the antiparticle-hole state.  Altogether, the quark field particle-anti-hole state is connected to the charge-conjugate antiparticle-hole state (or equivalently, to the quark field particle-antihole state itself), and the quark field antiparticle-hole state is connected to the charge-conjugate particle-antihole state (or equivalently, to the quark antiparticle-hole state).  
  
\begin{figure}
\includegraphics[scale=0.35]{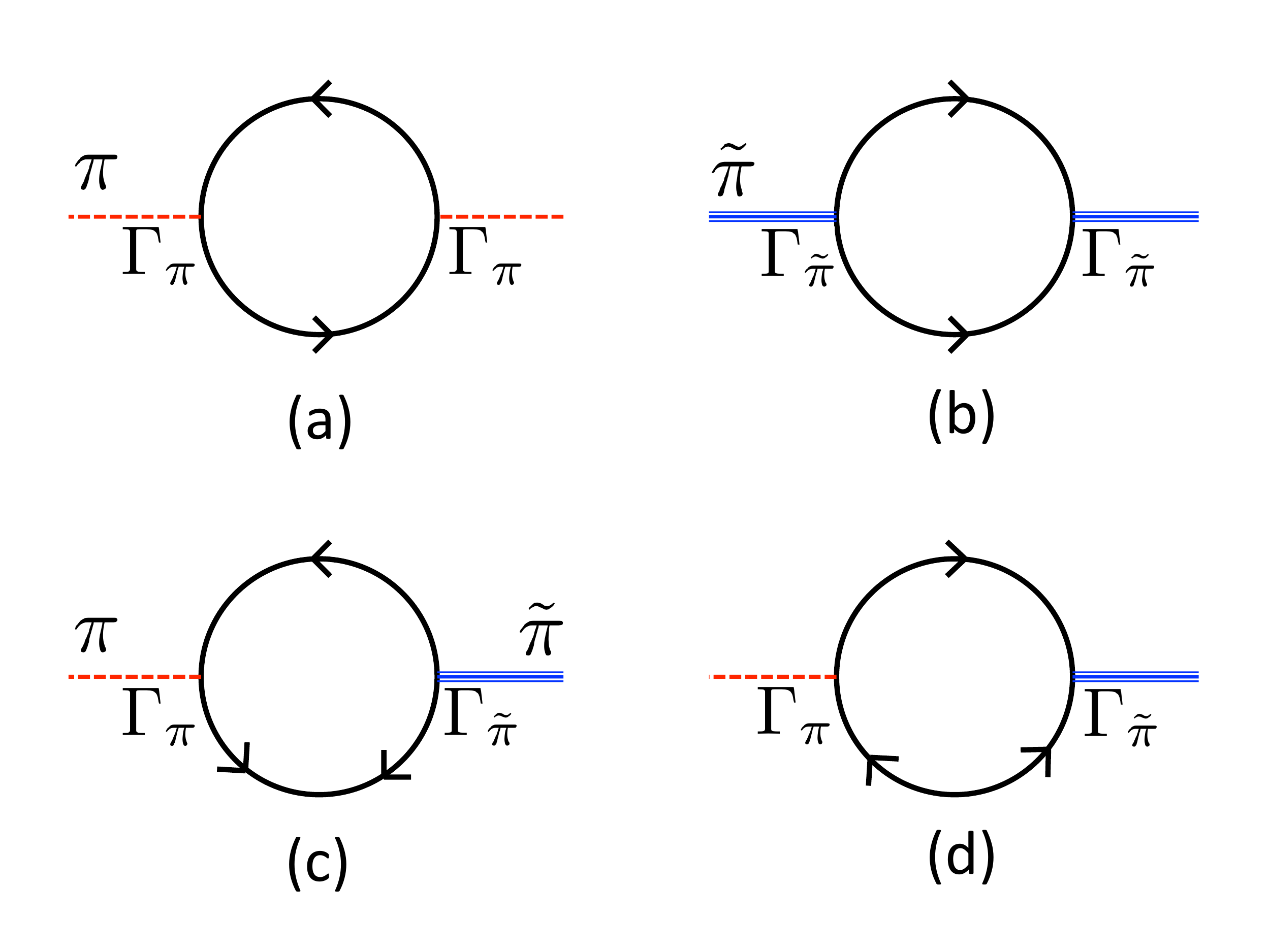}
\caption{\label{fig:bubbles}(Color online) Characteristic diagrams corresponding to the bubbles (\ref{eq:bubble_normalstate}).  The direct (a) $\pi$-$\pi$ and (b) $\pid$-$\pid$ bubbles correspond to $B_{\pi\pi}$ and $B_{dd}$, while the  $\pi$-$\pid$ mixing bubbles such as (c) and (d) correspond to $B_{\pi d}$. Due to the breaking of $U(1)_V$ by diquark pairing, quark number is not conserved.}
\end{figure}

  Furthermore, the mixing bubble $\Pi_{\pi d}$ can be further simplified by using the properties, Eq.~(\ref{eq:UV_identies}), of the coherence functions; we find
\begin{align}
B_{\pi d}(k_0^2) = -M\Delta\int_{\bm{p}}\sum_{j,\ell=\pm}\frac{\left(\epsilon_{j}-\epsilon_{\ell}\right)^{2}}{2\epsilon_{j}\epsilon_{\ell}\omega_{j}\omega_{\ell}}A_{j\ell}(k_{0}).
\end{align}
This bubble, connecting the chiral pion mode and diquark mode, is non-vanishing only in the coexistence region $M\neq 0$ and $\Delta \neq 0$.  The bubbles are summarized diagrammatically in Fig.~\ref{fig:bubbles}.

    In terms of the bubbles, the matrix $\Xi$ is given by $-\mathcal{D}_\theta^{-1} (k=0)$, as in Eq.~(\ref{eq:Pi}).   We find explicitly,  
\begin{align}
 & M^{2}\left(B_{\pi\pi}(0)-\frac{1}{2G}\right)=\Delta^{2}\left(B_{dd}(0)-\frac{1}{2H}\right)\nonumber \\
 & =-M\Delta B_{\pi d}(0) =-M^{2}\Delta^{2}a,
\label{eq:relation_M}
\end{align}
($a$ is given by Eq.~(\ref{eq:a})) confirming the expected form (\ref{eq:M_res}) of $\Xi$.

   From Eq.\,(\ref{eq:bubble_normalstate}) we calculate the ${\mathcal{Q}}$ matrix:
\begin{align}
\mathcal{Q}_{11} & =M^{2}\int_{\bm{p}}\sum_{j,\ell=\pm}\left(u_{j}v_{\ell}-v_{j}u_{\ell}\right)^{2}\left(1-\frac{M^{2}+\bm{p}^{2}}{\epsilon_{j}\epsilon_{\ell}}\right) W_{j\ell},\nonumber \\
\mathcal{Q}_{22} & =\Delta^{2}\int_{\bm{p}}\sum_{j,\ell=\pm}\left(v_{j}v_{\ell}+u_{j}u_{\ell}\right)^{2}\left(1-\frac{M^{2}-\bm{p}^{2}}{\epsilon_{j}\epsilon_{\ell}}\right) W_{j\ell},\nonumber \\
\mathcal{Q}_{12} & =-2M^{2}\Delta^{2}\int_{\bm{p}}\frac{1}{\omega_{+}\omega_{-}\left(\omega_{+}+\omega_{-}\right)^{3}}=\mathcal{Q}_{21},
\label{eq:F_normalstate}
\end{align}
where 
\begin{equation}
W_{j\ell}(\bm{p}) \equiv \frac1{(\omega_j(\bm{p}) + \omega_\ell(\bm{p}))^3}.
\end{equation}

The results (\ref{eq:M_res}) and (\ref{eq:F_normalstate}) show
that both $\Xi$ and the diagonal elements ${\mathcal{Q}}_{11}$ and ${\mathcal{Q}}_{22}$ are
logarithmically divergent (of order $\ln\Lambda/M$ or $\ln\Lambda/\Delta$),
while the off-diagonal elements ${\mathcal{Q}}_{12}$ are finite.  
 In the following, we drop the finite off-diagonal terms, following the standard prescription of considering only the ultraviolet-divergent pieces up to logarithmic accuracy of the bubble diagrams in effective bosonized theories (see e.g., \citep{Eguchi,Volkov1984,Ebert1983}). The dropped $Q_{12}$ terms would result in anomalous crossing terms $\sim\partial_\mu \hat{\theta}_\pi \partial^\mu \hat{\theta}_d$ which are absent in general parametrizations of pionic mode kinetic energies (up to second order in the pionic fields) in the literature, e.g.,\,\citep{hatsudaCollective}. 
 After this procedure, we identify the remaining diagonal elements of $\mathcal{Q}$ as the squared decay constants for the vacuum pion and the diquark-condensate pion:
\begin{equation}
f_{\pi}^{2}={\mathcal{Q}}_{11},\quad f_{\tilde{\pi}}^{2}={\mathcal{Q}}_{22};\label{eq:fpitoF}
\end{equation}
that is, $\mathcal{Q} = \mathcal{F}^2 = \mbox{diag}(f_\pi^2, f_{\tilde{\pi}}^2)$, where $\mathcal{F}=\mbox{diag}(f_\pi, f_{\tilde{\pi}})$ as in Eq.~(\ref{eq:pi_to_theta}).  Similarly dropping the finite off-diagonal terms of the velocity matrix $\mathcal{Q}_v$, we obtain ${\mathcal{Q}}_v = \mbox{diag}(v_\pi^2, v_{\tilde{\pi}}^2)\mathcal{Q}$, where the velocities are 
\begin{align}
v_\pi^2 = \mathcal{Q}_{v11} = f_\pi^{-2} \frac{\partial B_{\pi\pi} (0)}{\partial \bm{k}^2}, \nonumber \\
v_{\tilde{\pi}}^2 = \mathcal{Q}_{v22} = f_{\tilde{\pi}}^{-2} \frac{\partial B_{dd} (0)}{\partial \bm{k}^2}.
\end{align}

In terms of the pion fields $\pi(x)=f_{\pi}\hat{\theta}_\pi(x)$ and
$\tilde{\pi}(x)=f_{\tilde{\pi}}\hat{\theta}_{d}(x)$,
the effective Lagrangian density is now
\begin{align}
 & \frac{1}{2}\vec{\theta}\,^{T}\left({ - \mathcal{Q}}\partial_t^{2} + \mathcal{Q}_v \vec{\partial}^{\,2} -{\Xi}\right){\vec{\theta}}\nonumber \\
 & \equiv \frac{1}{2} \vec{\pi}^T ( - \partial_t^2 + \mbox{diag}(v_\pi^2, v_{\tilde{\pi}}^2)\vec{\partial}^{\,2} - \M) \vec{\pi},
\label{eq:thetatopiprop}
\end{align}
where $\vec{\pi}(x)\equiv (\pi(x),\,\tilde{\pi}(x))^{T} = \mathcal{F}\vec{\theta}(x)$.  The inverse propagator in Eq.~(\ref{eq:thetatopiprop}) is again diagonalized by Eq.~(\ref{eq:piGBtopi_1}), in terms of the NG mode $\pi_G$ and the massive mode $\pi_M$.
Furthermore, in terms of $\hat\theta_\pi$ and $\hat\theta_d$, we write
\begin{equation}
\pi_G=\frac{f_{\pi}^{2}\hat\theta_{\pi}+f_{\tilde{\pi}}^{2}\hat\theta_{d}}{\sqrt{f_{\pi}^{2}+{f}_{\tilde\pi}^{2}}}\equiv f_{G}\hat\theta_{G}\label{eq:piGBtotheta}
\end{equation}
where $\hat\theta_{G}$, the chiral NG
boson degree of freedom, is the fluctuation corresponding to the universal
axial $U(1)_{A}$ rotation of the whole system from the scalar state; such rotation corresponds to the simultaneous rotation of $\hat{\theta}_\pi$ and $\hat{\theta}_d$, therefore $\hat{\theta}_G = \hat{\theta}_\pi = \hat{\theta}_d$. As a result,
\begin{equation}
f_{G}^{2}=f_{\pi}^{2}+f_{\tilde{\pi}}^{2},\label{eq:fpirelation}
\end{equation}
thus relating the decay
constant of the NG boson $f_{G}$ to the decay constants  $f_{\pi}$ and $f_{\tilde{\pi}}$ for the corresponding chiral rotations of the $\langle \bar qq\rangle$ and  $\langle qq\rangle$  order parameters.  As Eq.~(\ref{eq:piGBtotheta}) shows $f_\pi^2$ and $f_{\pid}^2$ can be understood as the ``weight functions'' of $\pi$ and $\pid$ within the NG mode $\pi_G$. 

\begin{figure}
\includegraphics[scale=0.35]{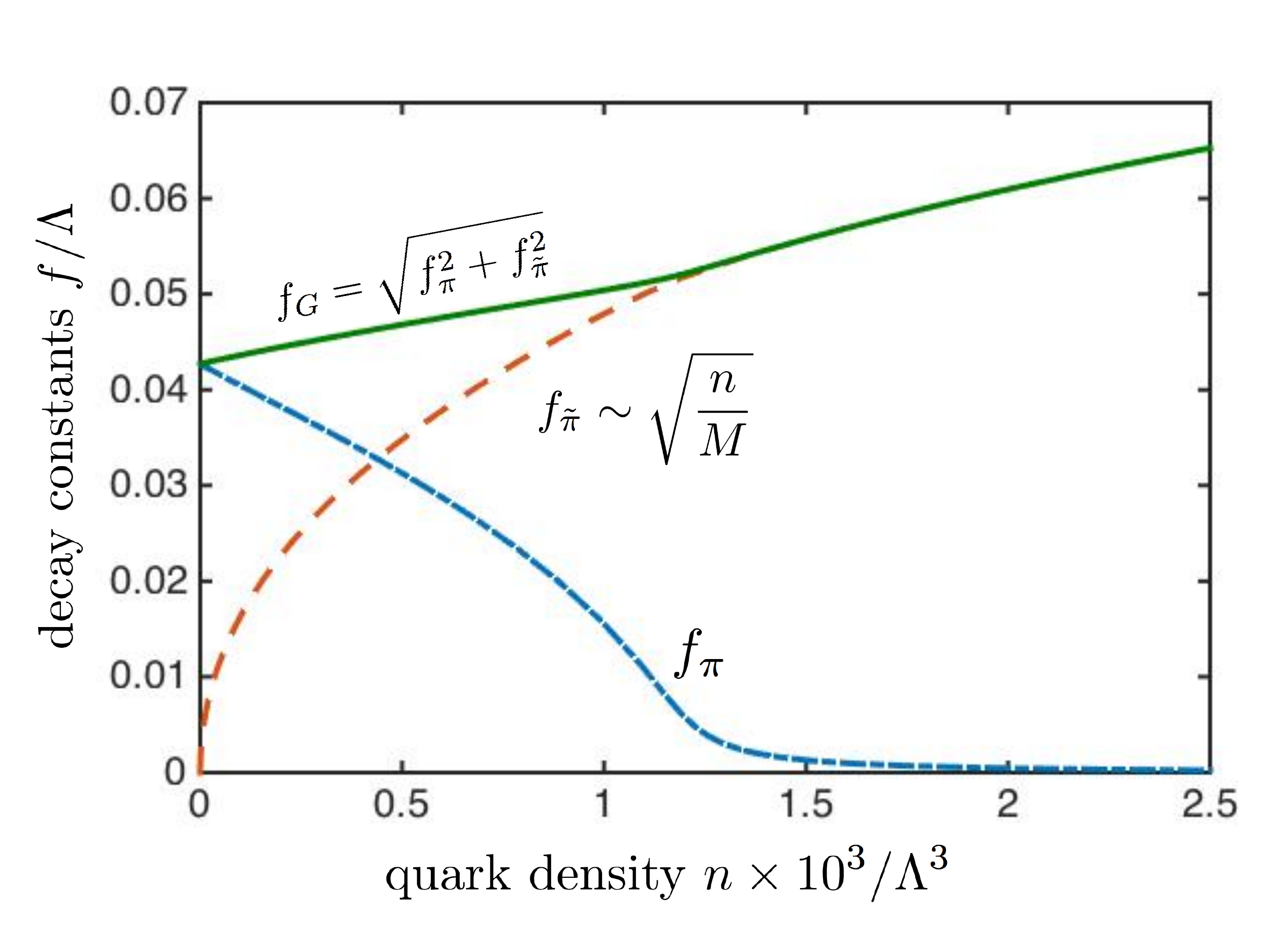}
\caption{\label{fig:decayf}(Color online) Decay constants $f_G$, $f_\pi$ and $f_{\tilde{\pi}}$ as functions of quark density $n$, with $G=11\Lambda^{-2}$ and $H=6\Lambda^{-2}$.}
\end{figure}

  The plot of $f_{G}$, $f_{\pi}$ and $f_{\tilde{\pi}}$ as functions of quark density in Fig.~\ref{fig:decayf} shows that the decay constant of the NG pion, $f_G$, always increases with increasing quark density, whereas $f_\pi$ decreases with density; the behavior of $f_\pi$ is in agreement with the prediction of in-medium chiral perturbation theory \cite{Goda2014} that to leading order in the density the pion decay constant decreases from its vacuum value linearly.\footnote{ Unlike NJL discussions of quark matter, reference \cite{Goda2014} discusses only a nucleon medium.  Although the vacuum cannot be described by deconfined NJL quark matter, the behavior of its chiral NG mode under modification of the density does connect qualitatively well with such nuclear matter models, a similarity suggesting that the transition from nuclear matter to high density quark matter could have continuous dynamic chiral symmetry breaking.   } 
    The different behavior of $f_{G}$ and $f_{\pi}$ arises from the presence of diquark pairing at all densities in our schematic model; even at low density, the BCS gap causes $f_G$ to increase with increasing density despite $\langle \bar{q} q\rangle$ (and thus $f_\pi$) gradually shrinking at the same time.\footnote{Realistically, the homogeneous diquark pairing described in the present model does not appear in the low density QCD phase diagram,  owing to the onset of confinement.} 
  
    The low density behavior of $f_{\tilde{\pi}}$ and $\Delta$ can be derived from the pairing gap equation (\ref{eq:GE_Delta_normalstate}) and the bubble results Eq.~(\ref{eq:F_normalstate}). Isolating the divergent part $1/\omega_+$ in the gap equation integral, one can show that in the limit $n \to 0$, the gap behaves like 
\begin{align}
\frac{\Delta}{\Lambda} \sim \frac{p_{F}}{M} e^{-\pi^2/HMp_{F}},
\label{eq:Delta_lown}
\end{align}
indicating that $\Delta/n$ goes to 0 as $n$ goes to 0.  Similarly, by isolating the divergent piece of the bubble integral in Eq.~(\ref{eq:F_normalstate}) in the $(j,\ell) = (+,+)$ part of the sum, one can show that in the limit $n\to 0$, $f_{\tilde{\pi}}\sim \sqrt{n/M}$, and $\Delta^2 / f_{\tilde{\pi}}^2 \sim \Delta^2 M/n \to 0$.

The decay constant $f_G$ can be equivalently parametrized as the vector transition amplitude from a state with one generalized pion to the vacuum via the time component of the axial current $J_A^\mu \equiv \bar{\psi} i\gamma^\mu \gamma^5 \psi/2$, in the same way as in the vacuum pion treatment \citep{buballa2005review} in NJL models:
\begin{align}
i f_G k^0 = & \langle 0 | J_A^0 | \pi_G\rangle =  \frac{1}{f_G} \langle 0 | J_A^0 | f_\pi \pi + f_{\pid} \pid  \rangle \nonumber \\
= & \frac{1}{f_G} i (f_\pi^2 + f_{\pid}^2) k^0,
\end{align}
again confirming Eq.~(\ref{eq:fpirelation}).  

\begin{figure}
\includegraphics[scale=0.35]{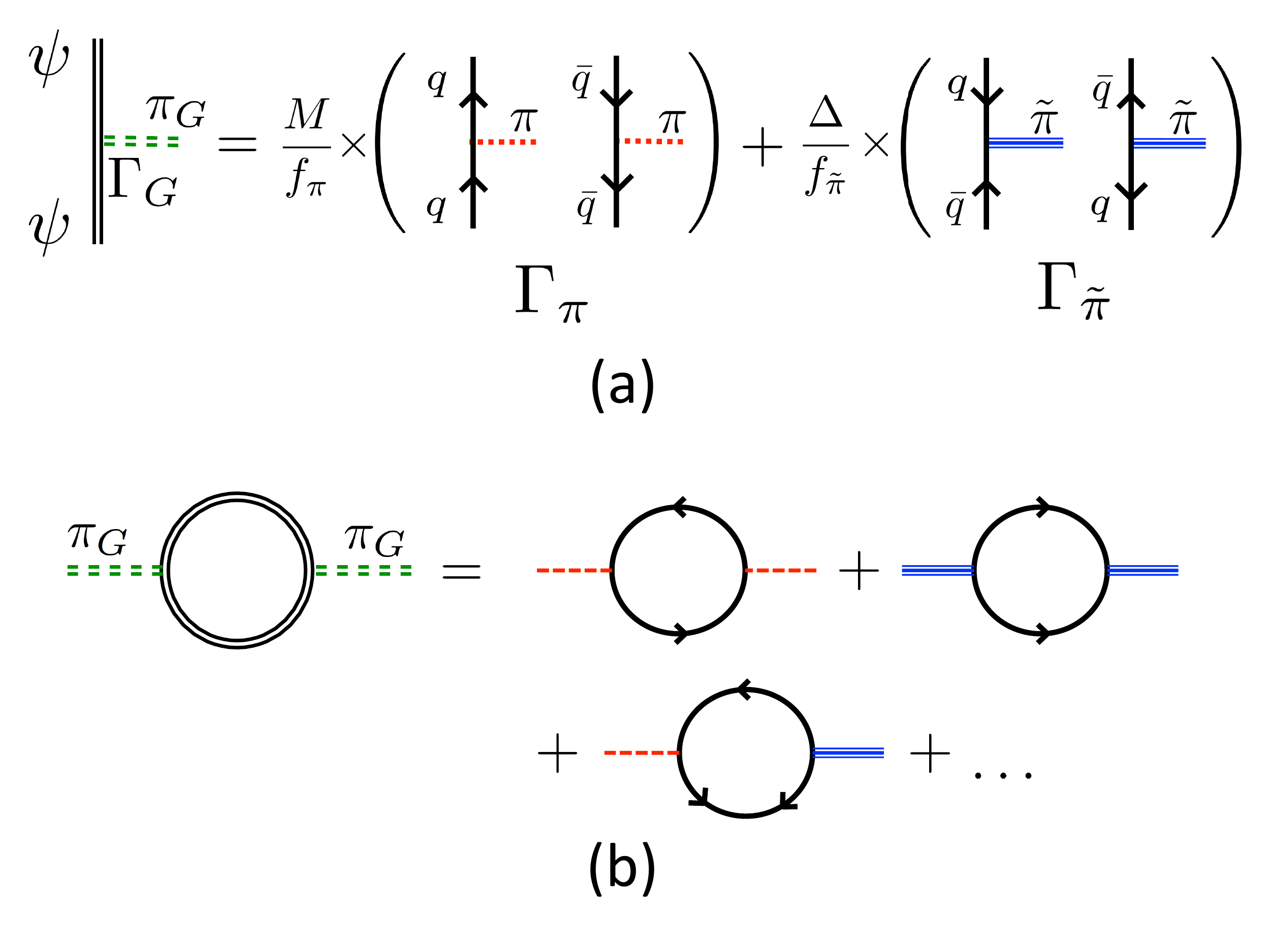}
\caption{\label{fig:interaction} (Color online) (a) Diagrammatic decomposition of quark-$\pi_G$ coupling $\Gamma_G$ into chiral  $\Gamma_\pi$ and diquark $\Gamma_{\pid}$ vertices. The (green) dashed double line represents the $\pi_G$ field.  The Nambu-Gor'kov field $\psi$ (black, double line) contains both the quark and charge-conjugate quark fields, thus including the quark field (black, solid, arrowed line) propagating in either time direction; $\Gamma_\pi$ is the coupling matrix between vacuum pion $\pi$ (red, dashed line) and the pseudoscalar $\bar{q}q$ quark sector, and $\Gamma_{\pid}$ is the coupling matrix between diquark-condensate pion $\pid$ (blue, double line) and the pseudoscalar $qq$ sector.   (b) Characteristic bubble diagrams contributing to the resulting self-energy of $\pi_G$ in the Nambu-Gor'kov formalism, including both direct bubbles, $B_{\pi\pi}$ and $B_{dd}$, and mixing bubbles, $B_{\pi d}$.   }
\end{figure}

    The density-dependent Nambu-Gor'kov interaction vertices coupling $\pi_G$ and $\pi_M$ to the Nambu-Gor'kov quark field $\psi$ can also be written in terms of the decay constants.  Using the perturbed quark inverse propagator with the bosonized fields in Eqs.~(\ref{eq:Sinv_pert}) and (\ref{eq:xdef}), and the transformation Eq.~(\ref{eq:piGBtopi_1}), we write the bosonized interaction as
\begin{align}
\mathcal{L}_{\subint} & =\bar{\psi}\left(\frac{M}{f_{\pi}}\Gamma_{\pi} \pi+\frac{\Delta}{f_{\pid}} \Gamma_{\pid} \pid \right)\psi \nonumber \\
& = \bar{\psi}\left(\Gamma_{G}\pi_{G}+\Gamma_{M}\pi_{M}\right)\psi,
\end{align}
where the interaction vertices,
\begin{align}
\Gamma_{G}(\mu) & \equiv \frac{1}{f_{G}}\left(M \Gamma_{\pi}+\Delta \Gamma_{\pid}\right), \nonumber \\
\Gamma_{M}(\mu) & \equiv \frac{1}{f_{G}}\left( \frac{f_{\pid}}{f_\pi} M \Gamma_\pi - \frac{f_\pi}{f_{\pid}} \Delta \Gamma_{\pid}\right),
\label{eq:piqqmatrix}
\end{align}
 are matrix functions of $\mu$, describing the coupling of $\pi_G$ and $\pi_M$ to the chiral $\bar{q}q$ and diquark pairing $qq$ sectors of the quark medium. Figure \ref{fig:interaction} represents the diagrammatical representation of the decomposition of $\Gamma_G$.  The coupling strengths to the chiral sector and the diquark sector are given by the weightings $M/f_G$ and $\Delta/f_G$; in the vacuum limit $\Delta/f_G = 0$, and the former simply reduces to $g_\pi$, the residue of the pion pole in the $\bar{q}q$-$\bar{q}q$ scattering T-matrix, related to $M$ and $f_\pi$ via the familiar Goldberger-Treimann relation $g_\pi = M/f_\pi$. 
  
  In more realistic NJL models with multiple flavors present,  possible asymmetric chiral and diquark pairings due to the heavy strange quark, and the Kobayahsi-Miskawa-'t Hooft six-quark instanton interaction \citep{KM1970, 't_Hooft_instanton, 't_Hooft_Symbreaking_BJanomalies} provide additional $\bar{q}q$-$qq$ mixing, with further modifications of $\Gamma_G$ and $\Gamma_M$.  We leave this topic to a future publication.

\subsection{\label{subsec:mode_mq}Finite bare quark mass $m_{q}\protect\neq0$}

We now turn on a finite but small bare
quark mass  $m_{q}$, explicitly breaking the chiral symmetry, to investigate its effect on the mass matrix $\M$. We
obtain the perturbed $\M$ by directly
taking second order derivatives of $\Omega$ with respect to $\theta_{\pi}$
and $\theta_{d}$, using Eq.\,(\ref{eq:FE_full}).  

    We first review the familiar $\sigma$-$\pi$ sector alone, where there is only one mode $\pi$ present. As
seen from the quasiparticle spectrum, $\omega_{\pm}(\bm{p})$, Eq.\,(\ref{eq:omega_full}),
$m_{q}$ slightly shifts $\sigma$, causing the system to
favor a negative value for $\sigma$ (whence the sign in 
the parametrization $\sigma=-M\cos\theta_{\pi}$). In the vacuum
scalar state, the eigenvalues expanded to leading order
in $m_q$ are :
\begin{eqnarray}
\omega_{\pm}(\bm{p}) & = & \left|\pm\left(\bm{p}^{2}+m_{q}^{2}+\sigma^{2}+\pi^{2}-2m_{q}\sigma\right)^{1/2}-\mu\right|\nonumber \\
 & = & \omega_{\pm}(\bm{p})_{m_{q}=0}\nonumber \\
 &  & +\left(\pm\frac{\mu}{\sqrt{\bm{p}^{2}+\sigma^{2}+\pi^{2}}}-1\right)\frac{\sigma m_{q}}{\omega_{\pm}(\bm{p})_{m_{q}=0}}\label{eq:omega_pi_mq}.
\end{eqnarray}
As a result, the free energy becomes
\begin{eqnarray}
\Omega & = & \Omega_{m_{q}=0}
 +\sigma m_{q}\sum_{\pm}\int_{\bm{p}}\frac{1}{\omega_{\pm}}\left(1\mp\frac{\mu}{\sqrt{\bm{p}^{2}+\sigma^{2}+\pi^{2}}}\right)\nonumber \\
 & \approx & \Omega_{m_{q}=0}+\frac{\sigma m_{q}}{2G},\label{eq:omega_Delta_mq}
\end{eqnarray}
where we have used the gap equation Eq.\,(\ref{eq:GE_m_normalstate}) in writing the second line, up to linear order in $m_q$.  With the
parametrization (\ref{eq:MF_para}), this term effectively adds a positive stiffness term for $\theta_{\pi}^{2}\sim\pi^{2}$, since $\sigma = -M(1-\theta_\pi^2/2 + \ldots)$.
We thus retrieve the GMOR result for the vacuum pion mass, 
\begin{align}
f_{\pi}^{2}m_{\pi}^{2}=\frac{M}{2G}m_{q}=-\langle\bar{q}q\rangle m_{q},
\label{eq:GMOR_vac}
\end{align}
to leading order linear in $m_{q}$.

     We also consider the pure BCS limit without the chiral $\sigma$-$\pi$ sector, setting $M=0$, and assuming zero phase difference $\phi$ between $\Delta_s$ and $\Delta_{ps}$.  The quasiparticle spectrum
becomes
\begin{equation}
\omega_{\pm}^{2}(\bm{p})=\bm{p}^{2}+\mu^{2}+\Delta^{2}\mp2\sqrt{\left(|\bm{p}|\mu\right)^{2}+m_{q}^{2}|\Delta_{ps}|^{2}}.\label{eq:omega_BCS_mq}
\end{equation}
The pseudoscalar diquark NG mode $\tilde{\pi}^2\sim\theta_{d}^{2}\sim|\Delta_{ps}|^{2}$
picks up a mass, given by
\begin{align}
f_{\tilde \pi}^{2}m_{\tilde \pi}^{2}=a \Delta^{2}m_{q}^{2},
\label{eq:GMOR_bcs}
\end{align}
as one sees from  the leading
order correction to $\Omega$, of order $m_{q}^{2}$, instead of $m_{q}$ for the $\pi$:
\begin{align}
\Omega  & =\Omega_{m_{q}=0}+\frac{1}{2} a m_{q}^{2}|\Delta_{ps}|^{2} + \mathcal{O}\left(m_{q}^{4}\right).
\label{eq:FE_BCS_mq}
\end{align}

Unlike in the $\sigma$-$\pi$ sector, the diquark mean fields $\Delta_{s}$ and $\Delta_{ps}$
are neither coupled directly nor offset by $m_{q}$ at the level of the mean field Lagrangian; instead, the diquark fields indirectly couple to $m_{q}$ via the mixing
term $|(m_{q}-\sigma)\Delta_{ps}-\pi\Delta_{s}|^{2}$. This term is 
the key to generating the mass of the NG mode in the BCS phase.

    The difference in the leading order dependence on $m_q$ of the GMOR relations in the vacuum phase and the high density BCS phase, which is also present in the more realistic $N_f=3$, $N_c=3$ case, can be understood as originating from the $U(1)_A$ axial symmetry.  Specifically, when one writes down a general Ginzburg-Landau effective Lagrangian in terms of the chiral and diquark condensates, the term of lowest non-zero order in $m_q$ and the diquark condensates that respects $U(1)_A$ symmetry is of order $m_q^2$ \citep{dtsonReversal, hatsudaCollective}.  As a result, at high density, where diquark pairing dominates, the chiral NG bosons should obey a GMOR relation $\sim m_q^2$.  A subtle complication in more realistic models is that the axial $U(1)_A$ symmetry is explicitly broken by quantum effects (the axial anomaly) at lower densities, which permits an additional mass term for the diquark condensates of order $m_q$.  In this case, the chiral NG bosons might still obey a GMOR relation $\sim m_q$ in leading order even with dominating diquark condensates at  moderate densities.  Nevertheless, it is known that at high density the axial anomaly is heavily suppressed \citep{Rapp2000, Schafer2002} greatly reducing such a $U(1)_A$-violating term; the GMOR relation is then restored to  $\sim m_q^2$ in leading order.\footnote{Diquark pairing is not the only known mechanism that can modify the meson mass GMOR relation.  The asymmetry in quark flavors could have a similar effect of inducing higher order GMOR relations, such as pions in an isospin-asymmetric medium \citep{Wojciech1995, Wojciech1997}.}

   Finally we calculate the perturbed mass of $\pi_G$ and $\pi_M$, in the intermediate density coexistence phase.  The two limits
considered above -- the pure $\sigma$-$\pi$ sector limit and the pure BCS limit -- indicate that we must keep effects of $m_{q}$ up to second order, and allow
fluctuations in both $\pi$ and $\Delta_{ps}$ -- achieved by small fluctuations of $\vec{\theta} = (\theta_\pi,\theta_d)^T$.
Expanding $\Omega$ in terms of $\vec{\theta}$
up to second order, we find
\begin{equation}
\Omega(\theta_{\pi},\theta_{d})=\Omega(0,0)+\frac{1}{2}{\vec{\theta}}^{\,\,T}{\Xi}(m_{q}){\vec{\theta}}+\ldots\,,\label{eq:FE_thetaqua}
\end{equation}
where (cf. Eq.~(\ref{eq:M_res})) 
\begin{align}
{\Xi}(m_{q}) = \left(\begin{array}{cc}
bMm_{q}+aM^{2}\Delta^{2} & -aM\Delta^{2}\left(M+m_{q}\right)\\
-aM\Delta^{2}\left(M+m_{q}\right) & a\left(M+m_{q}\right)^{2}\Delta^{2}
\end{array}\right);\nonumber\\ 
\label{eq:M_mq}
\end{align}
here
\begin{equation}
b=\int_{\bm{p}}\sum_{\pm}\frac{1}{\omega_{\pm}}\left(1-\frac{\mu}{\epsilon_{\pm}}\right),
\label{eq:b}
\end{equation}
the integral on the right side of the gap equation (\ref{eq:GE_m_normalstate}), is a function of $M$, $\Delta$, $\mu$, and $m_q$.
In the chiral limit $m_q=0$ in the chirally broken phase with $M\neq 0$, $b$
reduces to $1/2G$.
In terms of the mass matrix $\M = \mathcal{F}^{-1} \Xi \mathcal{F}^{-1}$ for the pion fields $\vec{\pi}$, we obtain the following matrix generalization of the GMOR relation encompassing both modes:
\begin{align}
\mathcal{F}\M \mathcal{F} & =  M^{2}a\Delta^{2}\left(\begin{array}{cc}
1 & -1\\
-1 & 1
\end{array}\right)+Mm_{q}\left(\begin{array}{cc}
b & -a\Delta^{2}\\
-a\Delta^{2} & 2a\Delta^{2}
\end{array}\right) \nonumber \\
 & \hspace{0.4cm} +a\Delta^{2}m_{q}^{2}\left(\begin{array}{cc}
0 & 0\\
0 & 1
\end{array}\right) \nonumber \\
 & \equiv  \Xi+Mm_{q}\Xi_{\roma{I}}+a\Delta^{2}m_{q}^{2}\Xi_{\roma{II}}.
\label{eq:GMOR}
\end{align}
Equation (\ref{eq:GMOR}) can be readily generalized to systems with more complex chiral order parameters than $\langle \bar{q} q \rangle$ and $\langle qq \rangle$. Despite appearances Eq.~(\ref{eq:GMOR}) is not actually a series expansion in $m_q$, since $a$, $b$, $f_\pi$, $f_{\tilde{\pi}}$, $\Delta$ and $M$ are themselves functions of $m_q$.  

 The structure of Eq.~(\ref{eq:GMOR}) clearly reflects the underlying physics.  The leading term $\Xi$ is a consequence of Goldstone's theorem, as argued before.  The perturbations to the stiffness matrix $\delta \Xi \equiv Mm_q\Xi_{\roma{I}} + a\Delta^2m_q^2 \Xi_{\roma{II}}$ contain combinations of order parameters that violate the $U(1)_A$ chiral symmetry, such as $~ \sigma |\Delta_s|^2$ and $\sigma$; they are results of $m_q$ explicitly breaking chiral symmetry. 

 \begin{figure}
\includegraphics[scale=0.35]{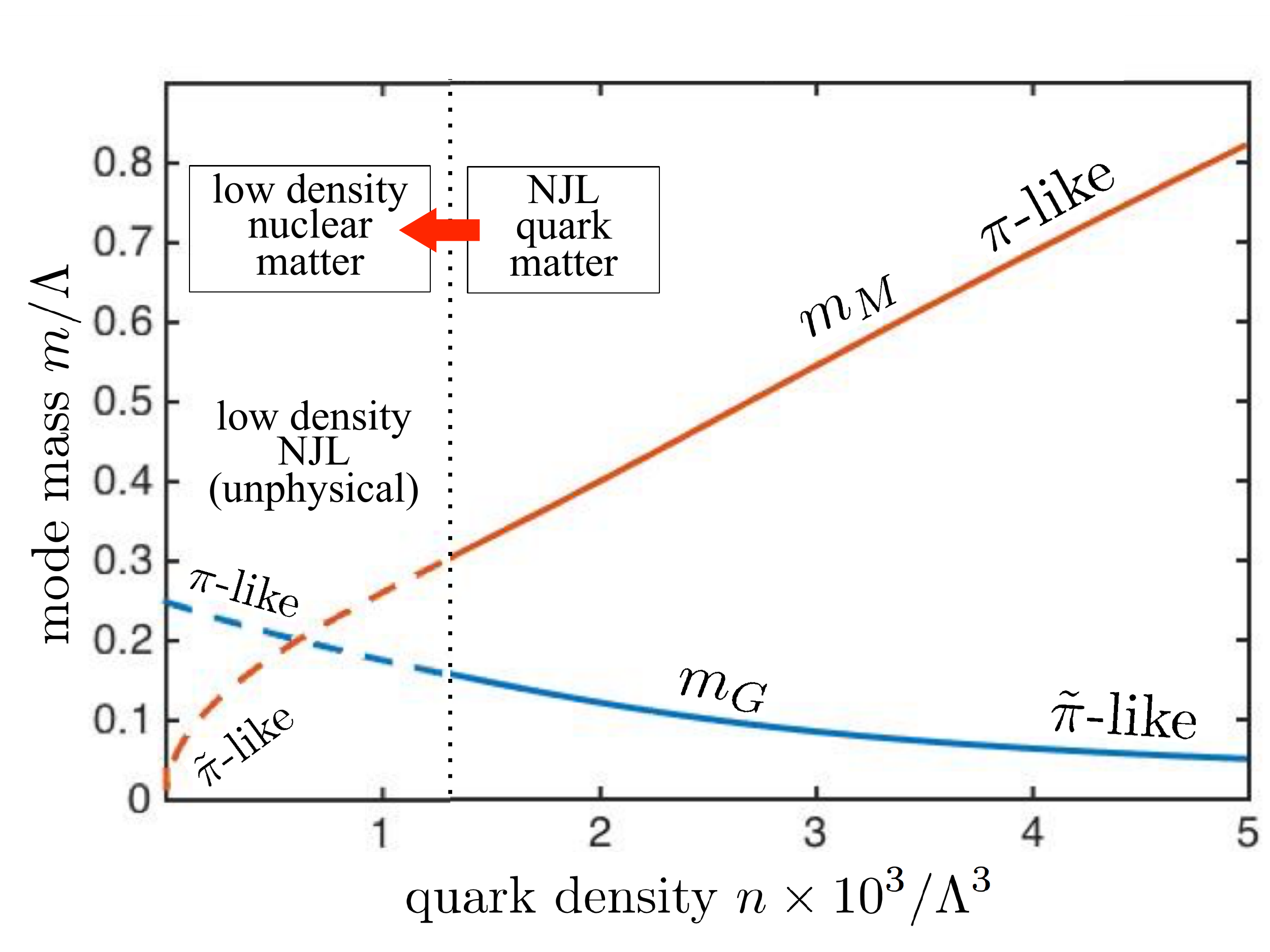}
\caption{\label{fig:mGmM}The perturbed masses of the NG mode, $m_G$, and of the heavy mode, $m_M$, as functions of quark density, $n$.  Here we take $G=11\Lambda^{-2}$, $H=6\Lambda^{-2}$ and $m_q = 0.01\Lambda$.  With decreasing density, $m_M$ rapidly decreases as the Fermi surface vanishes, eventually crossing the NG-mode mass $m_G$; this is an artifact of our simplified NJL model which does not take confinement into account.  Realistically this low density regime is instead described by nuclear matter; the boundary of the transition from quark matter to nuclear matter drawn in the plot is only illustrative. }
\end{figure}

  For non-zero $m_q$, to leading order in $\delta \Xi$ the perturbed squared masses are given by
\begin{align}
m_{G}^{2}  \approx & \, \frac{bMm_{q}+a\Delta^{2} m_{q}^{2}}{f_{G}^{2}},\nonumber \\
m_{M}^{2 } \approx & \,  aM^{2}\Delta^{2}\left(\frac{1}{f_{\pi}^{2}}+\frac{1}{f_{\tilde{\pi}}^{2}}\right)+Mm_{q}\left(\frac{bf_{\tilde{\pi}}^{2}}{f_{G}^{2}f_{\pi}^{2}}+\frac{2a\Delta^{2}}{f_{\tilde{\pi}}^{2}}\right) \nonumber \\
 & +m_{q}^{2}\frac{a\Delta^{2}f_{\pi}^{2}}{f_{G}^{2}f_{\tilde{\pi}}^{2}}.
\label{eq:m1m2p}
\end{align}
Figure~\ref{fig:mGmM} shows $m_G$ and $m_M$ as functions of the quark density $n$.  In the relatively high density BCS regime, $m_G$  decreases with increasing density as a consequence of the increasing BCS pairing $\langle qq\rangle$ taking on the role of chiral order parameter; $f_{\tilde{\pi}}$ increases while $f_\pi$ vanishes.  From the mixing, Eq.~(\ref{eq:piGBtopi_1}), one sees that the $\pi_G$ mode is mainly composed of $\tilde{\pi}$-like fluctuations, while the massive mode is mainly $\pi$-like, being heavy due to vanishing $\langle \bar{q}q\rangle$.   The NG-mode mass obeys the diquark-condensate pion GMOR relation (cf. Eq.~(\ref{eq:GMOR_bcs})):
\begin{align}
f_G^2 m_G^2 \approx am_q^2\Delta^2.
\end{align}
  
  At low density, the $\pi_G$ mode is primarily $\pi$-like, and one recovers the vacuum pion GMOR relation (cf. Eq.~(\ref{eq:GMOR_vac})):
\begin{align}
f_G^2 m_G^2 \approx b M m_q \approx \frac{M m_q}{2G} \approx - \langle\bar{q}q\rangle m_q
\end{align}
to leading order in $m_q$.  

  On the other hand, since the heavy mode $\pi_M$ is $\tilde{\pi}$-like at low density, $m_M$ vanishes in the limit $n\to 0$, crossing with $m_G$ in the process.  Such behavior is an artifact of the present schematic model: since diquark pairing is present at arbitrarily low densities in the model, the $\tilde{\pi}$-like mode, corresponding to chiral fluctuations of pairing amplitude $\langle qq\rangle$ mainly near the Fermi surface, the free energy cost goes to zero as the Fermi surface vanishes.  In the vacuum this mode is simply not present.
    
  The density at which $m_{M}$ crosses $m_G$ can be roughly estimated using Eq.~(\ref{eq:m1m2p}) and the fact that $\Delta \ll f_{\tilde{\pi}}$ at low density (see Eq.~(\ref{eq:Delta_lown}) and its comments) to show that when $m_G \sim m_M$, the decay constants are comparable with each other: $f_\pi \sim f_{\tilde{\pi}}$.  Since $f_{\tilde{\pi}}^2 \sim {n/M}$ at low density, $f_\pi \sim f_{\tilde{\pi}}$ implies $n \sim f_\pi^2 M$, a characteristic density scale for chiral symmetry breaking via $\langle \bar{q}q\rangle$.  Using values from realistic NJL models where the effective quark mass $M$ is $\sim 300 \,\mbox{MeV}$ and the experimental $f_\pi$ is $\sim$ 92 MeV, we find that $n$ is of order nuclear matter density, $n_0\approx 0.16 \,\mbox{fm}^{-3}$.  In this density regime, QCD confinement binds quarks into nucleons, and the homogeneous diquark pairing picture in the schematic model at these densities is no longer physical.  Nevertheless, the $\pi_G$ mode does obey the well-known vacuum pion GMOR relation in the low density limit, allowing this pionic mode to be smoothly interpolated between nuclear matter and  quark matter at high density, where  chiral symmetry remains broken  throughout.

\section{\label{sec:Outlook}Outlook}

   Having elaborated the construction and the density-dependent behavior of the generalized pion $\pi_G$, we briefly discuss several implications of the results obtained so far, and open questions for future research.

    \textit{1. Bose-Einstein condensation of the generalized pion $\pi_G$.} The existence of the light $\pi_G$ mode at all densitites, as detailed in Sec.\ref{subsec:Collective_modes}, raises the interesting possibility of the modes becoming Bose condensed.  Homogeneous condensates of the pionic NG modes have been considered, within NJL, in both the low density non-BCS (e.g., \citep{Anderson_piCnjl2}) and high density BCS (e.g., \citep{buballa_piCondensate_CFL}) limits.  In the present schematic model however, such condensation is trivial, since it merely corresponds to a global axial $U(1)_A$ rotation of the system from the scalar state.  In the chiral limit, the chiral symmetry is respected by the Lagrangian, and such rotation does not cost any free energy; the rotated system is energetically equivalent to the original scalar state.  With a finite $m_q$ breaking chiral symmetry, the scalar state is the unique ground state with the lowest free energy, since there are no forces driving condensation, and homogeneous pion condensates are unstable.

     In more realistic NJL models, however, where multiple flavors and charge neutrality are taken into account, several factors driving pionic condensation emerge.  For example, the mismatched Fermi surfaces of up and down quarks and an electric charge chemical potential translate directly into an effective chemical potential of the charged pions (see discussions of pion condensation in NJL models in \citep{Bedaque2001, Anderson_piCondensation_2Nf_chargeneutral, He2005, Ebert2006, Ebert2006a, Hao2007, Abuki2008a, Abuki2009}).   When the effective pion chemical potential overwhelms the pion mass, even homogeneous pion condensation can occur.  Furthermore, as the pions directly couple to the quarks in the pseudoscalar $\bar{q} q$ and $qq$ channels as discussed in Sec.~\ref{subsec:mode_decay},  more types of pionic condensates could be favored by the pion interacting with the quark matter medium at different densities, such as inhomogeneous meson condensates (e.g., \citep{Forbes_NJL_kaoncondensation, Gubina_Inhomo_picondensation11NJL}) or condensation into states with finite momenta.  Other exotic phases involving inhomogeneous chiral or diquark condensates (e.g., \citep{buballa_Inhomo_chiral_condensates, Alford_LOFF}) could also affect pion condensation.  We will discuss these possibilities in a future publication.
     
     \textit{2. Generalized meson mass ordering reversal. }  Reproducing the mass ordering reversal phenomenon as discussed in Ref.~\citep{dtsonReversal} again requires generalizing the present schematic model to three flavors and colors, with the strange quark heavier than the up and down quarks, and allowing for asymmetric pairing between the three flavors and colors due to mismatched Fermi surfaces at intermediate density.  The masses and decay constants of the generalized meson octet as functions of density can then be computed in the same way to study the the density-dependent meson mass spectrum throughout different phases, and how those mass curves depend on model parameters.  Such an analysis is required for further study of generalized meson condensation in realistic quark matter.
     
     \textit{3. Connection to nuclear matter pions.}  The interaction between vacuum pion mode $\pi$ and the quarks are the same as the nucleon-pion interaction in sigma model.  When diquark pairing is taken into account, the pion-quark interaction is modified into the density-dependent generalized $\Gamma_G$ vertex, which significantly reduces the generalized pion mass at higher density (see Fig.~\ref{fig:mGmM}).                 
  It is thus natural to ask whether nucleon-nucleon pairing at relatively high density (but still within the nuclear matter regime) would result in a similar modification to the generalized pion properties; as a consequence a one-to-one mapping between chirally broken nuclear matter to chirally broken quark matter in terms of generalized pions may be formed.  Such a continuity in chiral symmetry breaking would provide further insight into  possible continuity between nuclear matter and quark matter.
     
    \textit{4. Possible role of the $\hat{\phi}$ mode.} This mode, discussed in Sec.~\ref{subsec:Collective_modes}, could also play a role in a realistic phase diagram (a possibility that has not received attention in present NJL studies).  Although the $\hat{\phi}$ mode, not being a NG mode, is always massive, there may be density regions where its mass is significantly reduced.  This observation comes from the fact that the $\hat{\phi}$ mode corresponds to a relative phase oscillation between the scalar diquark condensate $\Delta_s$ and the pseudoscalar $\Delta_{ps}$.   Specifically, its stiffness term,
\begin{equation}
\frac{\partial^{2}\Omega}{\partial\sin^{2}\phi}=\frac{\Delta^{4}}{16}\sin^{2}2\theta_{d}\sum_{\pm}\int_{\bm{p}}\frac{1}{\omega_{\pm}^{3}}>0,
\label{eq:phi_mass}
\end{equation}
(calculated here, for simplicity, in the pure BCS limit with a finite $\theta_d$ chiral rotation from the scalar state in our model) can be made small if either the BCS gap $\Delta$ or the (homogeneous generalized pion condensation) $\theta_d$ is small.  The possible role of the $\phi$ mode in the low energy physics of dense quark matter and its coupling to the pseudoscalar $\Delta_{ps}$ fluctuations and thus its coupling to the generalized pion will be explored in a future study as well.

\section*{Acknowledgments}  
We are grateful to Tetsuo Hatsuda for his careful critique of this paper.   This research was supported in part by NSF Grant PHY1305891.

\end{document}